\documentclass[fleqn,usenatbib]{mnras}

\usepackage{newtxtext,newtxmath}

\usepackage[T1]{fontenc}

\usepackage{gensymb}

\usepackage[flushleft]{threeparttable}

\DeclareRobustCommand{\VAN}[3]{#2}
\let\VANthebibliography\thebibliography
\def\thebibliography{\DeclareRobustCommand{\VAN}[3]{##3}\VANthebibliography}

\usepackage{graphicx}	
\usepackage{amsmath}	

\usepackage{tikz,xcolor,hyperref}

\definecolor{lime}{HTML}{A6CE39}
\DeclareRobustCommand{\orcidicon}{%
    \begin{tikzpicture}
    \draw[lime, fill=lime] (0,0) 
    circle [radius=0.16] 
    node[white] {{\fontfamily{qag}\selectfont \tiny ID}};
    \draw[white, fill=white] (-0.0625,0.095) 
    circle [radius=0.007];
    \end{tikzpicture}
    \hspace{-2mm}
}
\newcommand{\orcidYSun}{\href{https://orcid.org/0000-0001-6561-9443}{\orcidicon}}
\newcommand{\orcidGHLee}{\href{https://orcid.org/0000-0001-9501-1252}{\orcidicon}}
\newcommand{\orcidAZabludoff}{\href{https://orcid.org/0000-0001-6047-8469}{\orcidicon}}
\newcommand{\orcidDFrench}{\href{https://orcid.org/0000-0002-4235-7337}{\orcidicon}}
\newcommand{\orcidJHelton}{\href{https://orcid.org/0000-0003-4337-6211}{\orcidicon}}
\newcommand{\orcidNKerrison}{\href{https://orcid.org/0000-0002-8977-1009}{\orcidicon}}
\newcommand{\orcidYJYang}{\href{https://orcid.org/0000-0003-3078-2763}{\orcidicon}}

\newcommand{\hda}{{\rm H}\delta_{\rm A}}
\newcommand{\OIII}{\mbox{O\,\textsc{iii}}} 
\newcommand{\NIII}{\mbox{N\,\textsc{iii}}} 
\newcommand{\NII}{\mbox{N\,\textsc{ii}}} 
\newcommand{\SII}{\mbox{S\,\textsc{ii}}} 
\newcommand{\OI}{\mbox{O\,\textsc{i}}} 
\newcommand{\MgII}{\mbox{Mg\,\textsc{ii}}} 
\newcommand{\NaI}{\mbox{Na\,\textsc{i}}}
\newcommand{\nad}{\mbox{Na\,\textsc{D}}} 
\newcommand{\hei}{\mbox{He\,\textsc{i}}}
\newcommand{\heii}{\mbox{He\,\textsc{ii}}}
\newcommand{\ewexc}{{\rm EW_{NaD,exc}}}
\newcommand{\ewtot}{{\rm EW_{NaD,tot}}}
\newcommand{\lgmstar}{{\rm log}(M_{*}/M_{\odot})}

\newcommand{\mstar}{M_{*}/M_{\odot}}
\newcommand{\kms}{\rm km~s^{-1}}

\newcommand{\angstrom}{\mbox{\normalfont\AA}}

\title[Evolution of Gas Flows in Galaxies]{Evolution of Gas Flows along the Starburst to Post-Starburst to Quiescent Galaxy Sequence}

\author[Y. Sun et al.]{
Yang Sun,\orcidYSun $^{1}$\thanks{E-mail: sunyang@arizona.edu}
Gwang-Ho Lee,\orcidGHLee $^{1,2}$
Ann I. Zabludoff,\orcidAZabludoff $^{1}$
K. Decker French,\orcidDFrench $^{3}$
Jakob M. Helton,\orcidJHelton $^{1}$
\newauthor{
Nicole A. Kerrison,\orcidNKerrison $^{1}$
Christy A. Tremonti,$^{4}$
and Yujin Yang \orcidYJYang $^{2}$
}
\\
$^{1}$Department of Astronomy and Steward Observatory, University of Arizona, 933 North Cherry Avenue, Tucson, AZ 85721, USA\\
$^{2}$Korea Astronomy and Space Science Institute, 776 Daedeokdae-to, Yuseong-gu, Daejeon, 305-348, Republic of Korea\\
$^{3}$Department of Astronomy, University of Illinois, 1002 West Green Street, Urbana, IL 61801, USA \\
$^{4}$Department of Astronomy, University of Wisconsin–Madison, 475 North Charter Street, Madison, WI 53703, USA
}

\date{Accepted 2024 January 31. Received 2024 January 26; in original form 2023 September 30}

\pubyear{2024}

\begin{document}
\label{firstpage}
\pagerange{\pageref{firstpage}--\pageref{lastpage}}
\maketitle

\begin{abstract}
We measure velocity offsets in the $\NaI$ $\lambda\lambda5890, 5896$ ($\nad$) interstellar medium absorption lines to track how neutral galactic winds change as their host galaxies evolve.  
Our sample of $\sim$80,000 SDSS spectra at $0.010 < z < 0.325$ includes starburst, post-starburst, and quiescent galaxies, forming an evolutionary
sequence of declining star formation rate (SFR). 
We detect bulk flows across this sequence, 
mostly at higher host stellar masses ($\lgmstar)>10$).
Along this sequence,
the fraction of outflows
decreases ($76\pm2$\% to $65\pm4$\% to a 3$\sigma$ upper limit of $34$\%), and
the mean velocity offset changes from outflowing to inflowing ($-84.6\pm5.9$ to $-71.6\pm11.4$ to $76.6\pm2.3\,\kms$).
Even within the post-starburst sample, wind speed decreases with time elapsed since the starburst ended. These results reveal
that outflows diminish as 
galaxies age. For post-starbursts, there is evidence for an AGN contribution, especially to the speediest outflows: 
1) SFR declines faster in time than outflow velocity, a decoupling arguing against massive stellar feedback;
2) of the few outflows strong enough to escape the interstellar medium (9/105), three of the four hosts with measured emission lines are Seyfert galaxies. For disky starburst galaxies, however, the trends suggest flows out of the stellar disk plane (with outflow 1/2-opening angle
$> 45\degree$) instead of from the nucleus: the wind velocity decreases as the disk becomes more edge-on, and the outflow fraction, constant at $\sim$90$\%$ for disk inclinations $i<45\degree$,
steadily decreases from $\sim$90$\%$ to 20$\%$
for $i>45\degree$.
\end{abstract}




\section{Introduction}\label{sec.intro}

In the local Universe, observational surveys suggest there are two main types of galaxies: star-forming and quiescent. The current understanding of this bimodality is that star-forming galaxies will eventually stop forming stars and evolve to the quiescent phase. Transitioning galaxies evolving between star-forming and quiescent are found in the “green valley” of the galaxy color-magnitude diagram \citep{2004ApJ...600..681B}. Among transitioning galaxies are post-starburst galaxies, which experienced a burst of star formation that ended in the past $\sim1~\mathrm{Gyr}$ \citep{Dressler_spectroscopy_1983,couch_spectroscopic_1987}. Evidence for this star formation history comes from their optical spectra, which show strong Balmer absorption lines that indicate a substantial population of A-type stars and thus a recent starburst but lack the nebular emission line signatures of current star formation. Even though star formation must decline along the evolutionary sequence from starburst to post-starburst to quiescent, the reasons why are still unknown.

One likely mechanism for this quenching is feedback. For current cosmological simulations of galaxy formation and evolution, feedback is required to reproduce observed galaxy baryonic mass functions and star formation efficiency \citep[][and references therein]{keres_galaxies_2009,hopkins_galaxies_2014,somerville_physical_2015,naab_theoretical_2017}. Massive stars, supernovae, and active galactic nuclei (AGN) may be sources of negative feedback through thermal or kinetic energy transferred to the interstellar medium (ISM) in galaxies. If negative feedback processes remove the cold gas in winds (see \citealt{veilleux_galactic_2005, king_powerful_2015,veilleux_cool_2020} for a comprehensive galactic wind review), star formation could rapidly end due to the removal of the fuel for star formation. 

Previous work has found observational evidence of outflows in low redshift galaxies, generally focusing on AGN-host galaxies \citep[e.g.,][]{crenshaw_intrinsic_1999,nesvadba_evidence_2008,rubin_low-ionization_2011,harrison_kiloparsec-scale_2014,cicone_massive_2014,kang_unraveling_2018,baron_discovering_2019} or on pure star-forming or starburst galaxies \citep[e.g.,][]{heckman_nature_1990,heckman_absorption-line_2000,rupke_outflows_2005,rupke_outflows_2005-1,chen_absorption-line_2010,rubin_evidence_2014,zschaechner_spatially_2018,perrotta_physical_2021}. For post-starburst galaxies, outflows have been detected in hosts with
\citep{baron_multiphase_2021} and without AGN \citep{coil_outflowing_2011}. The evolution of such outflows along the starburst to post-starburst to quiescent sequence, however, remains unprobed.

Outflows are detected in different gas phases, e.g., molecular gas \citep{cicone_massive_2014,fluetsch_cold_2018}, neutral gas \citep{chen_absorption-line_2010,bae_independence_2018,concas_two-faces_2019}, or ionized gas \citep{harrison_kiloparsec-scale_2014,concas_light_2017,bae_independence_2018}. The complex connections between multi-phase outflows and star formation or AGN activity have been well studied over the past decades, but a unified picture is elusive. For instance, \citet{concas_light_2017} analyzed a complete spectroscopic sample of galaxies in the SDSS DR7 survey and found that the flux percentage of the outflow component of the ionized gas outflows tracer [$\OIII$] $\lambda$5007 are higher in low-ionization nuclear emission-line regions (LINERs) and Seyfert galaxies than in pure star-forming galaxies, suggesting that such winds are powered by the nucleus.
On the other hand, \citet{chen_absorption-line_2010} systematically analyzed the stacked $\nad$ lines of star-forming galaxies in the SDSS DR7 survey, finding that significant ISM $\nad$ absorption lines (equivalent widths $\text{EW}>0.8~\angstrom$) are only prevalent in disk galaxies with high star formation rate (SFR), SFR per unit area ($\Sigma_{\text{SFR}}$), or stellar mass ($M_{\star}$). 

In their multi-phase outflow study of post-starburst galaxies, \citet{baron_multiphase_2021}
used both model-fitting and Machine Learning methods to select post-starburst galaxies with strong recent starbursts ($H\delta~\text{EW}>5~\angstrom$), AGN features (AGN-like narrow line flux ratios), and ionized outflows (multiple broad emission lines).
They found 144 eligible post-starburst galaxy candidates, $40\%$ of which were detected with $\nad$-traced neutral gas outflows. 
While they found a connection between the detection of $\nad$ and star-formation (SF) luminosity, the correlations of the neutral outflow velocity, mass outflow rate, and kinetic power with SF luminosity or AGN luminosity were not significant. The ionized outflow velocity is correlated significantly with both AGN luminosity and SF luminosity but has a stronger correlation with AGN luminosity.

The detection of neutral gas outflows is correlated with the disk inclination in starbursting galaxies.
\citet{heckman_absorption-line_2000} found $\sim$70\% of such outflows are detected in disks with inclinations ($i$) of less than $60 \degree$. Later works similarly found that most of the neutral gas outflows detected in absorption are in galaxies with low inclination
\citep{chen_absorption-line_2010,rubin_evidence_2014,roberts-borsani_prevalence_2018}. \citet{bae_independence_2018} confirmed the inclination dependence of neutral outflow velocity and velocity dispersion in both pure star-forming and AGN-host galaxies, 
interpreting this result as support for a 
perpendicular outflow geometry that 
could only arise from star formation-driven winds.
\citet{concas_two-faces_2019} observed that the velocities of neutral gas outflows decrease with inclination for their starburst galaxy sample.

Even though many studies have identified connections between SFR and wind properties, none have tested how winds evolve along the galaxy evolutionary sequence where star formation is declining. Based on previous findings that higher outflow detection rates and larger $\nad$ EWs prefer higher SFR and $M_{\star}$ host galaxies, we expect that there should be a decline in winds over this sequence. Therefore, in this paper, we build a sequence of declining star formation, from starburst to post-starburst to quiescent galaxies in the local Universe ($0.010 < z < 0.325$). We then explore the evolution of neutral gas winds along that sequence.
Within our post-starburst sample, we know even more: the time elapsed since the starburst ended
\citep{french_clocking_2018} (hereafter F18).
These ``post-burst ages'' allow us to
trace the wind evolution directly over time, from
roughly 0 to 2 Gyr after the end of the burst, with an age resolution of about $\pm$ 50 Myr.

We use the spectroscopic SDSS DR12 survey \citep{alam_eleventh_2015} and analyze the $\NaI$ $\lambda\lambda5890, 5896$ line to trace the bulk motions of the neutral gas in the starburst, post-starburst, and quiescent samples. We test how the detection of winds and the wind velocities change along the whole sequence and with post-burst age. We further investigate whether wind properties are related to host galaxy properties such as stellar mass and disk inclination. Finally, we consider the potential sources of the winds.

This paper is organized as follows: we describe the selection of our three galaxy samples in Section~\ref{sec.samp}. Our method of subtracting the stellar component from each SDSS spectrum, fitting the profile of the residual (ISM)$\nad$ absorption line, and deriving the direction (inflow vs. outflow) and line-of-sight velocity of the flow is described in Section~\ref{sec:anal}. We present the main results in Section~\ref{sec.results} and discuss them further in Section~\ref{sec.discussion}. Finally, we summarize our findings in Section~\ref{sec.concl}.

Throughout this paper, we assume a standard $\Lambda$CDM universe whose cosmological parameters are $\mathrm{H_0} =
70~\mathrm{km~s^{-1}~Mpc^{-1}}$, $\Omega_{\Lambda} = 0.7$, and $\Omega_{\mathrm{m}} = 0.3$.

\section{Sample Selection and Data}\label{sec.samp}

\begin{figure*}
	\includegraphics[width=\textwidth]{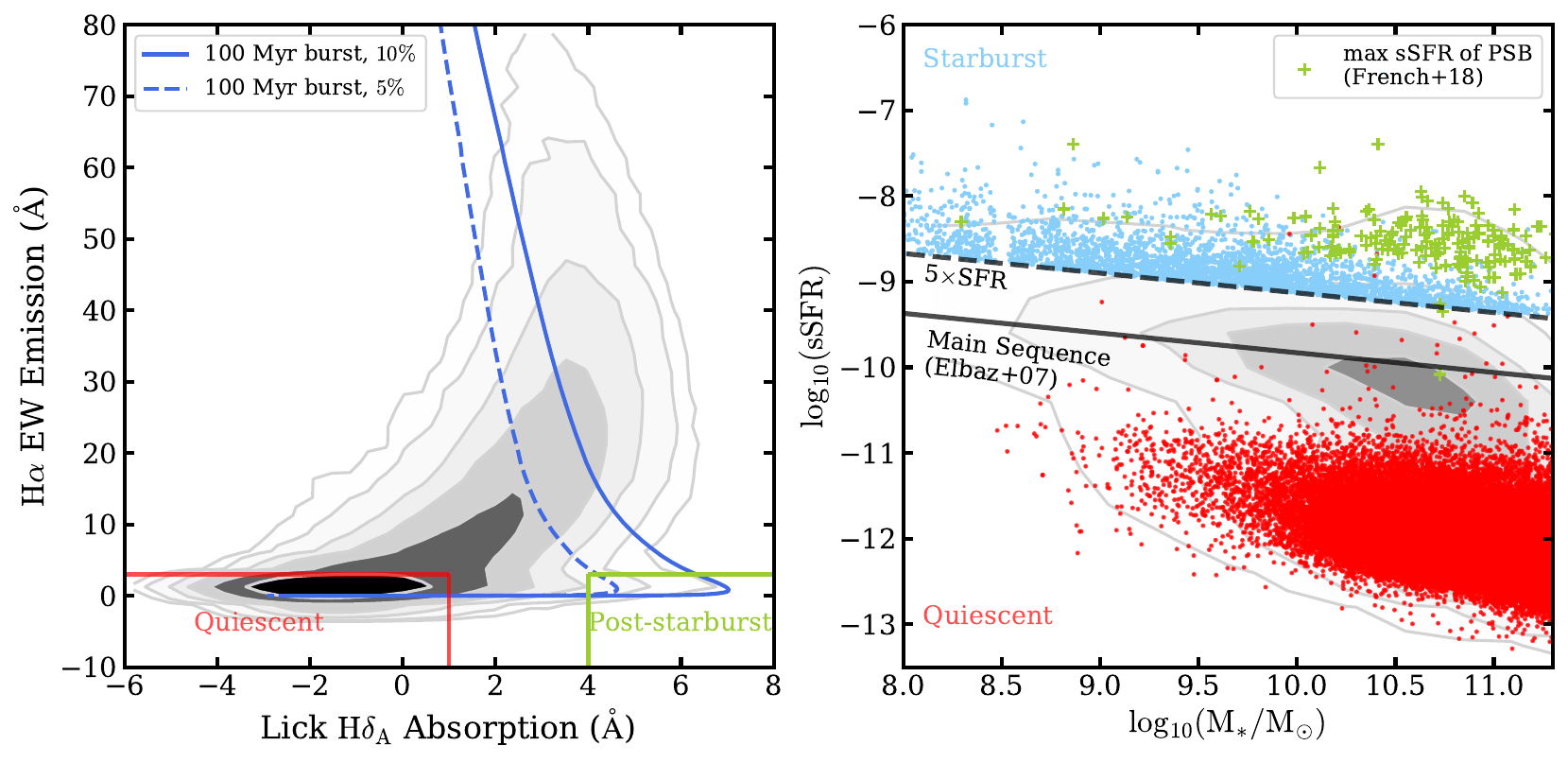}
    \caption{Definition of starburst, post-starburst, and quiescent galaxy samples. Left: Lick $\hda$ absorption index vs. H$\alpha$ emission equivalent width (EW) distribution for our parent sample of 593,807 galaxies at $\mathrm{0.010 < z < 0.325}$ drawn from the SDSS DR12 survey. The bottom-right region outlined in green represents post-starburst galaxy selection criteria used in \citet{french_clocking_2018}: H$\alpha$ EW$<3~\angstrom$ and $\hda-\delta(\hda)>4~\angstrom$, where $\delta(\hda)$ is the error in the Lick index. Quiescent galaxies are located in the bottom-left region delineated in red: $\hda+\delta(\hda)<1~\angstrom$ and H$\alpha$ EW$<3~\angstrom$. The dashed and solid blue lines show two example tracks of a $100\,\mathrm{Myr}$-long burst with a burst mass fraction of 5\% and 10\%, respectively, and are adapted from \citet{french_clocking_2018}. Right: Stellar masses ($M_*$) vs. specific SFRs (sSFRs) of 
    SDSS galaxies from the MPA-JHU catalogs (grey contours). The solid line is the “main sequence” from \citet{elbaz_reversal_2007}, while the dashed line indicates the $\mathrm{5\times SFR}$ level above which we select our starburst sample (blue points). The gap around $\lgmstar=8.5$ without any starburst galaxies arises artificially. 
    There are few post-starburst galaxies at stellar masses less than 9.5$M_{\odot}$
    from which to define the starburst progenitor distribution. In particular, for the sparsely-sampled range of post-starburst burst mass fractions, there are no corresponding starburst galaxies in our sample at $\sim$8.5$M_{\odot}$.
    Green crosses represent the maximum sSFRs during the bursts of post-starburst galaxies from \citet{french_clocking_2018}.
    Red points are the quiescent galaxies selected from the Lick $\hda$ absorption index vs. H$\alpha$ emission EW diagram (red box in the left panel).
    Based on the selection from these two panels, we obtain 3306, 516, and 72251 galaxies for the starburst, post-starburst, and quiescent samples, respectively.}
    \label{Fig:sample_select}
\end{figure*}
This study is based on three different galaxy samples: a post-starburst galaxy sample and two comparison
samples of starburst and quiescent galaxies.  The starburst sample is chosen to be consistent with the evolutionary precursors of the post-starburst sample. The quiescent sample is selected to be consistent with evolving from galaxies like those in the post-starburst sample. 

\subsection{Post-Starburst Galaxy Sample}\label{sec:PSBsamp}
F18 
drew 532 post-starburst 
galaxies  
from 
the spectroscopic galaxy catalog of the Sloan Digital Sky Survey 
Data Release 8 \citep[DR8,][]{aihara_eighth_2011}, 
using galaxy properties from 
the MPA-JHU Catalogs \citep{kauffmann_stellar_2003,brinchmann_physical_2004,tremonti_origin_2004}.
From the plane of 
the Lick H$\delta$ absorption line index ($\hda$, {\tt\string LICK\underline{ }HD\underline{ }A}) versus 
the H$\alpha$ emission line equivalent width
({\tt\string H\underline{ }ALPHA\underline{ }EQW}) (the left panel of Figure~\ref{Fig:sample_select}),
F18 identified the post-starburst sample from the lower right corner. They
selected for 
low H$\alpha~\text{EW}<3~\angstrom$ in emission, 
representing a lack of current star formation, 
and strong Balmer absorption line index 
$\hda-\sigma(\hda) > 4~\angstrom$, 
where $\sigma(\hda)$ is the measurement error, 
representing a large population of A-type stars
and thus 
a substantial burst of star formation 
in the past $\sim$1-1.5 Gyr.

Through a detailed modeling of the star formation 
histories (SFHs) of the post-starburst galaxies, 
F18 parameterized each SFH with the post-burst age, 
the fraction of stellar mass produced in the burst ($\mathrm{m_{burst}}$), 
and the burst duration. 
In their age-dating technique, 
they used Lick indices measured from the SDSS spectra. 
They excluded galaxies with signal-to-noise (S/N) less than 
10 pixel$^{-1}$ in the continuum 
to avoid unreliable measurements. 
The best constraints on the age-dating parameters require ultraviolet (UV) photometry. 
By matching with \emph{Galaxy Evolution Explorer} \citep[\textit{GALEX},][]{martin_galaxy_2005} far-UV (FUV) and near-UV (NUV) data, they finalized a sample of 532 post-starburst galaxies 
with $>3\sigma$ detections in both the NUV and FUV bands. 

After F18 selected their post-starburst sample, the MPA-JHU Catalogs were updated, and some line measurements were revised. As a result, nine of the original F18 galaxies 
no longer satisfy the post-starburst 
$\hda$-H$\alpha$
selection criteria, and another galaxy has an unreliable global stellar mass measurement ({\tt\string LGM\underline{ }TOT\underline{ }P50}). We also find the H$\alpha$ flux measurements of six galaxies are bad ({\tt\string H\underline{ }ALPHA\underline{ }CHISQ}$=$0).
Throughout this paper, after removing those 16 galaxies, 
we use the remaining 516 post-starburst galaxies from F18. 
Their redshift range is $0.010<z<0.325$.

Measuring current SFRs for post-starburst galaxies is complicated by the uncertainties due to dust geometry, the long timescale of star formation traced by some indicators, and contamination from possible AGN (see recent review in \citealt{french_evolution_2021}). H$\alpha$ is a relatively good SFR indicator in the post-starburst phase, because it traces a short star formation timescale (i.e., that associated with short-lived massive stars) and dust attenuation and AGN effects can be modeled and removed. 

Here we convert the H$\alpha$ fluxes from the MPA-JHU catalogs to SFR following \citet{french_discovery_2015} and \citet{li_evolution_2019}\footnote{We do not use the SFR measurements provided by the MPA-JHU catalogs, because, for 
non-star forming galaxies on the BPT diagram \citep{baldwin_classification_1981, veilleux_spectral_1987}, including most post-starburst galaxies, 
they are derived from the D4000 index
and the relation between D4000 and specific SFR (sSFR). 
D4000-traced SFRs may not be calibrated accurately for post-starburst galaxies, given that the D4000-sSFR relation is sensitive to the star formation history.}. 
We use the $L_{\mathrm{H \alpha}}$-SFR relation from \citet{kennicutt_past_1994}. We estimate the amount of internal dust extinction from the observed Balmer decrement, H$\alpha$/H$\beta$. Assuming that the hydrogen nebular emission follows Case B recombination, the intrinsic Balmer flux ratio $(\mathrm{H\alpha/H\beta})_0=2.86$ for $T_e=10^4$ K. We adopt the reddening curve of \citet{odonnell_r_1994}. When the H$\beta$ line flux is negative in the MPA-JHU catalogs, we assume the mean value of $E(B-V)$ from the other post-starburst galaxies. The mean attenuation is $A_V=0.92$ mag. 

Following the methodology from \citet{wild_timing_2010}, we correct for a potential underlying AGN contribution to the H$\alpha$ flux and thus to the derived SFR. We calculate the emission-line ratios [$\OIII$]/H$\beta$ and [$\NII$]/H$\alpha$ of our post-starburst galaxies, place them on the BPT diagram, and determine the likely AGN contributions to their H$\alpha$ luminosity. For 230 post-starbursts with [$\OIII$], [$\NII$], and H$\alpha$ signal-to-noise ratios ($S/N$) $<$ 3, we do not have a reliable AGN correction, so we do not measure their SFRs. We do not set a $S/N$ constraint on the H$\beta$ line, as it is not usually well-detected in our post-starburst galaxies. For the case of negative H$\beta$ line flux, we use its 1$\sigma$ flux upper limit to determine the corresponding 1$\sigma$ upper limit for the AGN correction to the SFR. 
When a negative correction factor is derived, if the corresponding 1$\sigma$ upper limit is positive, we use this upper limit to determine the SFR 1$\sigma$ upper limit; otherwise, we do not measure the SFR.

We then correct for the SDSS fiber aperture using the ratio of the global SFR to the SFR within the SDSS fiber ({\tt\string SFR\underline{ }TOT\underline{ }P50}/{\tt\string SFR\underline{ }FIB\underline{ }P50}) from the MPA-JHU catalogs. In the end, we obtain SFR measurements for 171 of the 516 post-starburst galaxies, including 79 post-starburst galaxies for which only SFR upper limit measurements are available. The bottom panel of Figure~\ref{Fig:SFR_sequence} shows that the SFRs of our post-starburst galaxies decline over time. We discuss the timescale of this decline in Section~\ref{sec.discuss_vwind_age}.

\subsection{Comparison Samples}\label{sec:ComparSamp}

We build a broader evolutionary sequence of declining star formation by constructing samples of likely post-starburst progenitors (starburst galaxies) and descendants (quiescent galaxies).
We start with a parent sample of 593,807 galaxies drawn from the SDSS DR12 survey \citep{alam_eleventh_2015}. These galaxies lie within the same redshift range as our post-starburst galaxies and have properties and spectral line flux measurements from the MPA-JHU catalogs.
All galaxies are selected to have continuum $S/N\geq$10.

\subsubsection{Starburst Galaxy Sample}\label{sec:SBGSamp}
As the progenitors of post-starburst galaxies, starburst galaxies should have strong on-going star formation by definition. Because the post-starburst sample from F18 was selected against strong AGN, we also require their progenitors to be without strong AGN features. 
While there are SDSS starburst galaxies with powerful AGN that could also evolve into galaxies like our post-starburst galaxies, we exclude them here to avoid having to decouple the timescale of AGN decline from that of the SFR. 

We first select an appropriate parent sample of 157,018 star \emph{forming} galaxies without strong AGN, 
using \citet{kewley_host_2006}'s classification scheme:
\begin{equation}
{\rm log([\OIII]/H\beta)<0.61/log([\NII]/H\alpha)-0.05])+1.3},
\end{equation}
with $S/N$ $\geq 3$ in the H$\alpha$,
H$\beta$, [$\OIII$], and [$\NII$] emission line fluxes. Any remaining galaxies that are classified as quasars by the SDSS pipeline
({\tt\string SpecClass=QSO}) or have invalid flux values around the $\nad$ region are excluded as well. 

To select which of these star-forming galaxies have SFRs characteristic of starburst galaxies, we consider the relation between 
H$\alpha$-based specific SFRs (sSFRs) and $M_*$ from the MPA-JHU catalogs (Figure~\ref{Fig:sample_select}, right panel). The solid line indicates the star forming main sequence from \citet{elbaz_reversal_2007} \footnote{log(sSFR) = -0.23$\mathrm{log(M}_{*}/{\rm M_{\odot}})-7.53$.}.
The 2-4$\times$SFR levels are often used to distinguish 
between the main sequence and starburst galaxies \citep{elbaz_goods_2011,rodighiero_lesser_2011,sargent_regularity_2014}.
There are 3723 galaxies located above the $5\times$SFR level 
(dashed line);
we define these as starburst galaxies and further justify that choice below.

To ensure that the starburst sample is the progenitor of the post-starburst sample, we consider the growth of the stellar mass based on the mass fraction of new stars created in the starburst $\mathrm{m_{burst}}$.
For each post-starburst galaxy, we determine the mass range of its potential progenitors as $[\mstar(1-\mathrm{m_{burst}}), \mstar]$ and choose all starburst galaxies within it. F18 found some post-starburst galaxies have poorly constrained burst mass fractions ($\mathrm{m_{burst}}>0.5$ and with large uncertainties). Therefore, we set these to 0.5, 
the 
$1\sigma$ upper limit on the average burst mass fraction for the better-determined values.
In the end, we select 3,306 galaxies as possible starburst progenitors of the post-starburst sample. The right panel of Figure \ref{Fig:sample_select} confirms that these starburst galaxies have 
sSFRs comparable to the maximum sSFRs of our post-starburst galaxies during their starburst phase (F18).

\subsubsection{Quiescent Galaxy Sample}\label{sec:QSCSamp}

Quiescent galaxies are selected to have $\hda+\delta(\hda)<1~\angstrom$ 
and H$\alpha$ EW$<3~\angstrom$, as shown in the left panel 
of Figure~\ref{Fig:sample_select}, with the intention of selecting galaxies with no current and recent star formation. 
These quiescent galaxies follow the example evolutionary tracks of post-starburst galaxies with different burst durations and burst mass fractions provided by F18 (adapted as blue lines in the left panel of Figure~\ref{Fig:sample_select}). These selection criteria result in 260,526 galaxies from the parent sample. Again, we remove galaxies with significant emission lines to exclude ongoing SF or AGN activity. In this step, 
we remove 176,125 galaxies with 
S/N$\geq$3 in at least one of the emission lines 
H$\beta$, [$\OIII$] $\lambda5007$, H$\alpha$, [$\NII$] $\lambda$6584, 
and [$\SII$] $\lambda\lambda6717,6731$. Also, 1,208 quiescent galaxies with incomplete spectra around the $\nad$ region are excluded. Our quiescent sample is finalized at 72,251 galaxies. 

Under the simple assumption that post-starburst galaxies evolve with their current residual SFRs (typically $\sim$0.1 $M_{\odot}~{\rm yr}^{-1}$) throughout the post-starburst phase, $10^8~M_{\odot}$ mass growth is expected over 1 Gyr. More generously, a maximum mass growth of $10^9~M_{\odot}$ can be expected over a Hubble time. Considering these mass growth bounds, we exclude 10,942 quiescent galaxies with stellar masses less than the minimum stellar mass ($M_* < 10^{8.29}~M_{\odot}$) of our post-starburst sample or greater than the maximum allowed stellar mass. Our quiescent sample is finalized at 72,251 galaxies. 

\begin{figure}
	\includegraphics[width=\columnwidth]{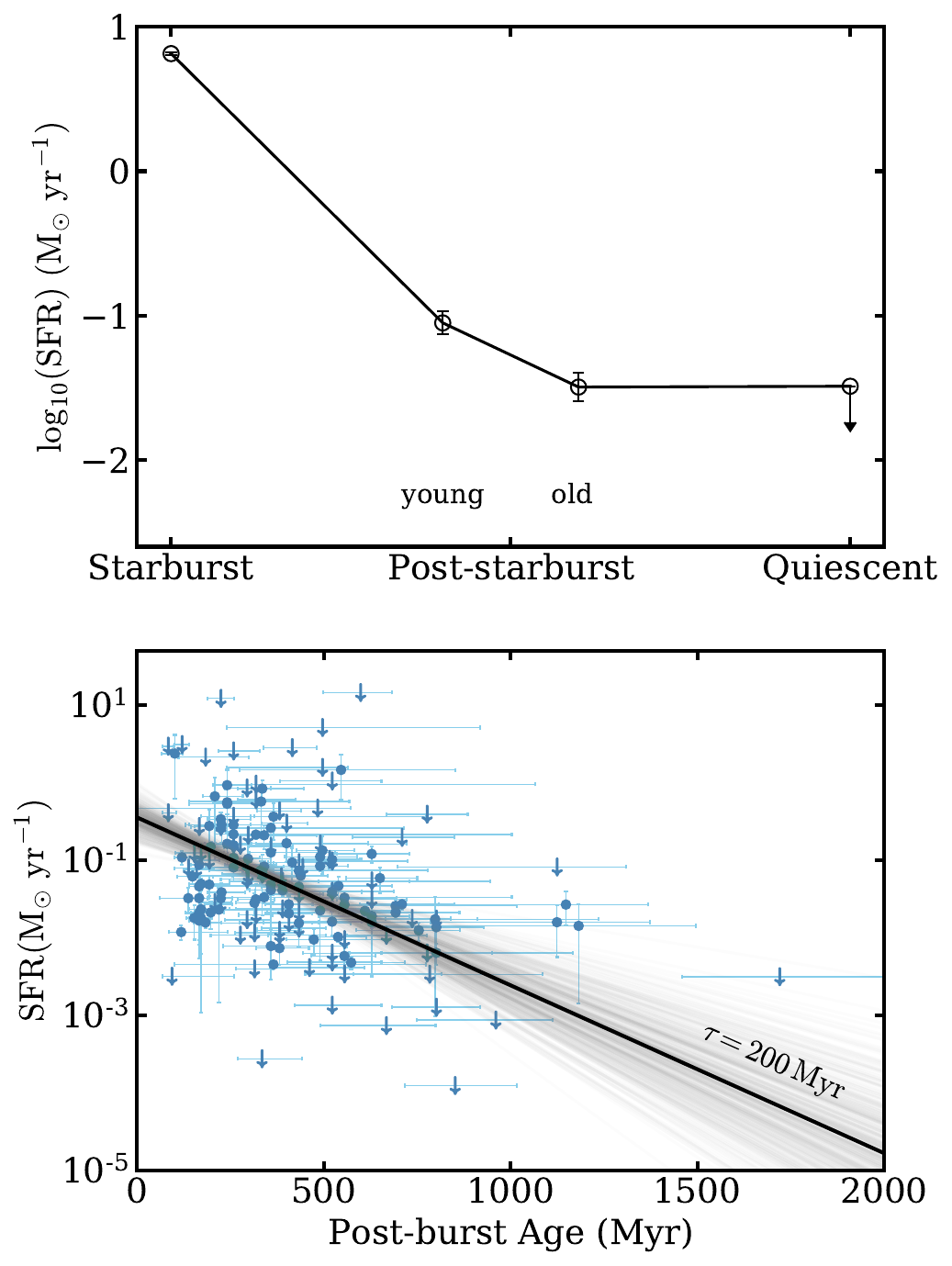}
    \caption{\textbf{Top}: $\mathrm{H\alpha}$-based SFR declines along the evolutionary sequence, from starburst to young post-starburst to old post-starburst to quiescent galaxies. The mean SFR of quiescents is an upper limit. \textbf{Bottom}: The SFR of post-starburst galaxies declines with their post-burst age. Arrows represent galaxies that have only $\mathrm{H\alpha}$-based SFR upper limits. The best-fit slope of the post-starburst age - $\ln(\mathrm{SFR})$ relation gives an exponential decline timescale of 200 $\pm 50$ Myr. These results confirm that SFR declines along the evolutionary sequence that we have constructed.}
    \label{Fig:SFR_sequence}
\end{figure}

For consistency with the starburst and post-starburst samples, we use the ${\mathrm{H \alpha}}$ luminosity to estimate SFRs for the quiescent galaxies, rather than the D4000-based SFRs provided by the MPA-JHU catalogs. Our selection for low levels of star formation in the quiescents requires low ${\mathrm{H \alpha}}$ values. Furthermore, we only measure SFR for those quiescents where ${\mathrm{H \alpha}}$ has $S/N > 1$. We additionally require that all BPT emission lines for the quiescents have
$S/N < 3$, which prevents us from correcting for
any AGN contribution.  Due to these restrictions, the SFRs for the quiescents are upper limits. In total, we obtain SFR measurements for 36,718 of 72,251 quiescent galaxies. The mean SFR of the quiescent sample (top panel of Figure~\ref{Fig:SFR_sequence}) is thus a rough upper limit.

To summarize, by following the above steps, we select starburst, post-starburst, and quiescent galaxy samples to build an evolutionary sequence of declining star formation. The top panel of Figure~\ref{Fig:SFR_sequence} shows the mean global SFR of starbursts ($\log_{10}(\mathrm{SFR})=0.82\pm0.01~M_{\odot}~\mathrm{yr^{-1}}$), young ($-1.05\pm0.08~M_{\odot}~\mathrm{yr^{-1}}$) and old ($-1.49\pm0.10~M_{\odot}~\mathrm{yr^{-1}}$) post-starbursts (young and old post-starbursts will be defined in \S\ref{sec.wind_prop_sequence}), and quiescents ($<-1.49~\mathrm{yr^{-1}}$). It confirms that the global SFR is declining along the sequence that we have constructed. Similarly, the bottom panel of  Figure~\ref{Fig:SFR_sequence} shows that the global SFR declines with post-burst age for the post-starburst sample.

\section{Analysis}\label{sec:anal} 

\begin{figure*}
	\includegraphics[width=0.98\textwidth]{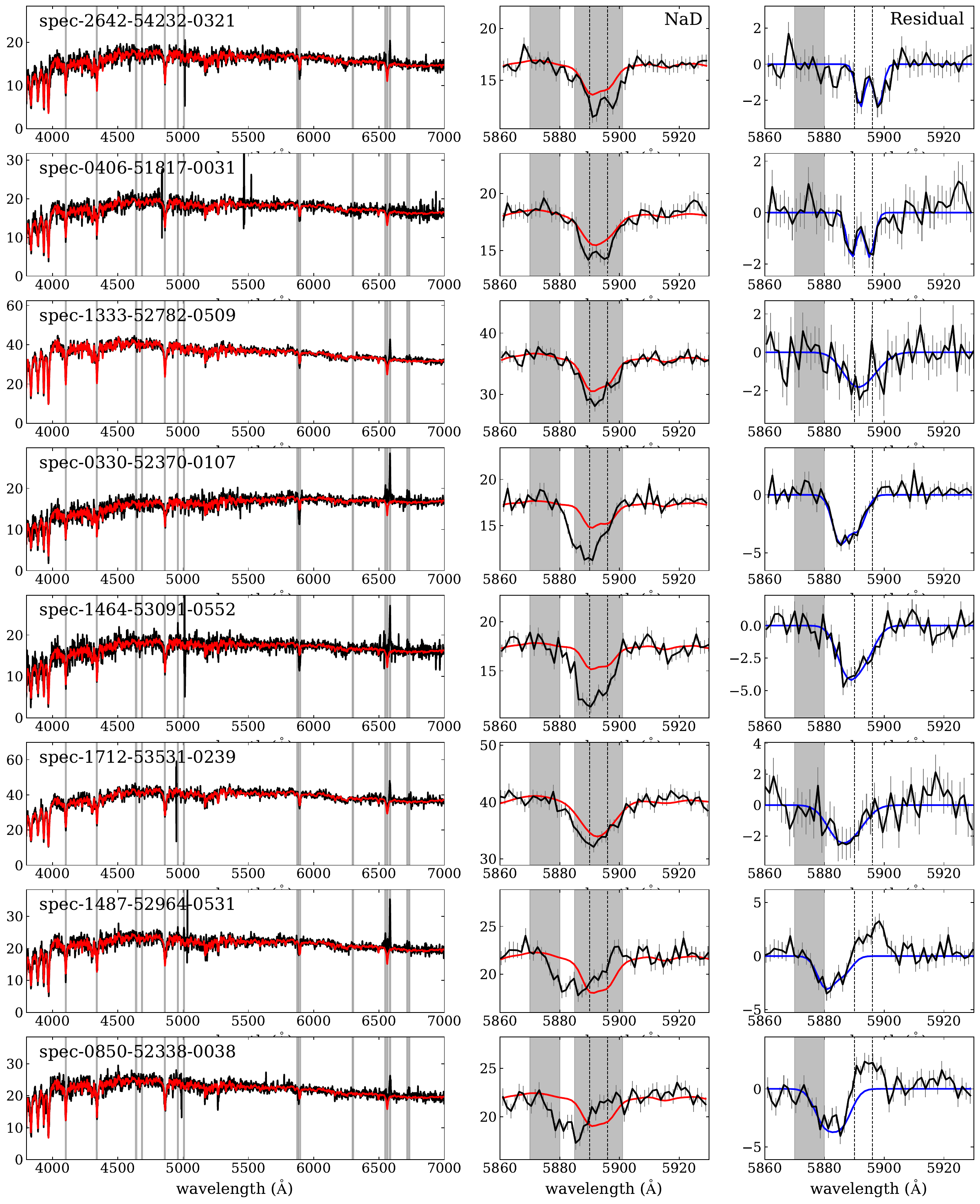}
    \caption{Examples of our spectral fitting for post-starburst galaxies. Left: SDSS spectra in the rest-frame (black) and their best-fit stellar continuum model obtained using pPXF (red).  Middle: Zoom-in around $\nad$ 5890, 5896$~\angstrom$ (dashed lines) spectra (black) with the best-fit stellar continuum models (red). Right: $\nad$ residual spectra (black), representing the ISM, and their best-fit double Gaussian profile (blue). Errorbars show the original flux uncertainties from the SDSS spectra. The masked regions are shown as gray areas. When fitting the residual $\nad$ spectra, the region around $\heii~\lambda~5875$ is still masked. The last two rows show two galaxies that have a P-Cygni profile in the $\nad$ residual that is composed of a blueshifted absorption component and a redshift emission component.}
    \label{Fig:PSB_sample}
\end{figure*}

\begin{figure*}
	\includegraphics[width=0.98\textwidth]{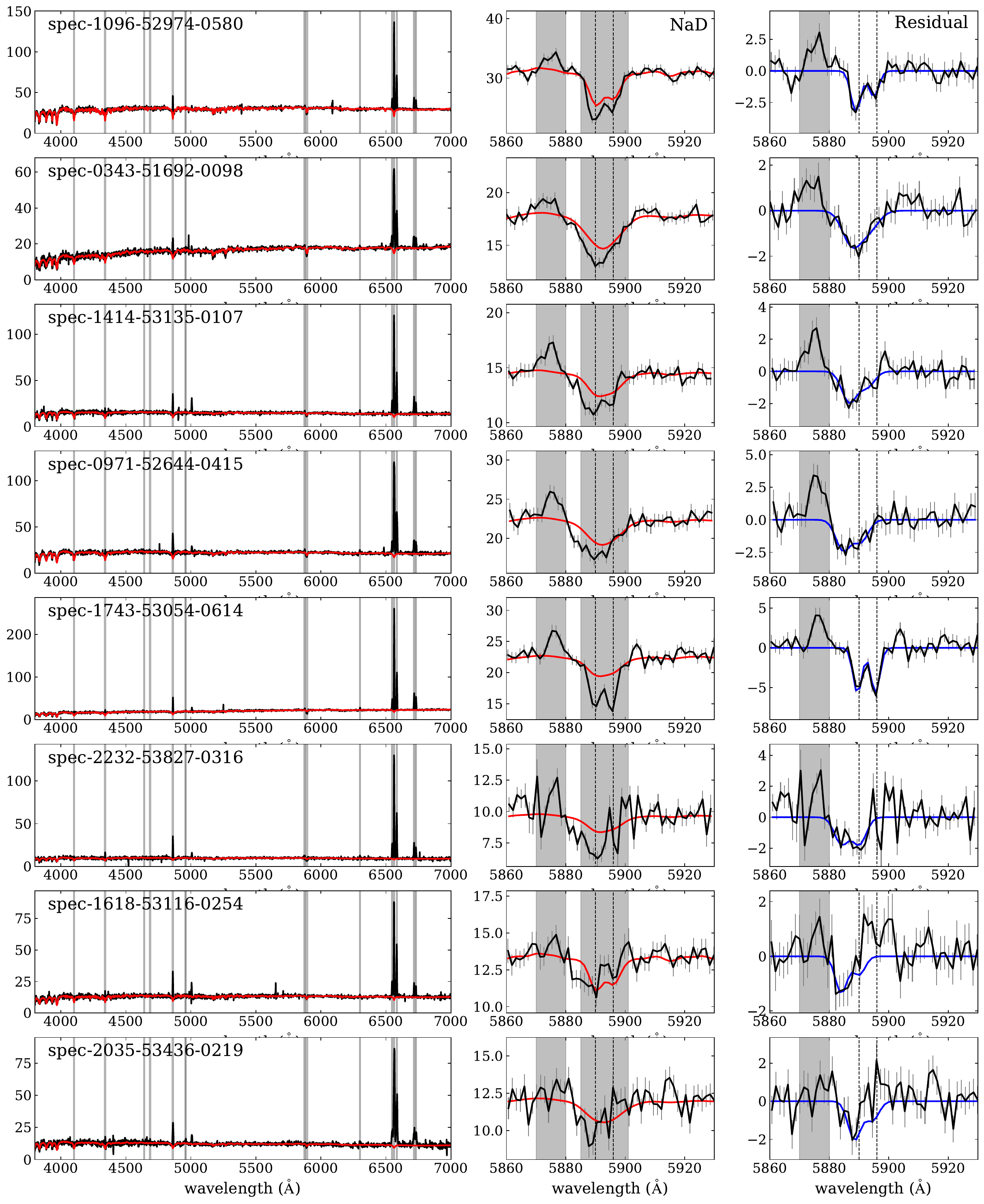}
    \caption{Same as Figure~\ref{Fig:PSB_sample}, but for our starburst sample.}
    \label{Fig:SBG_sample}
\end{figure*}

\begin{figure*}
	\includegraphics[width=0.98\textwidth]{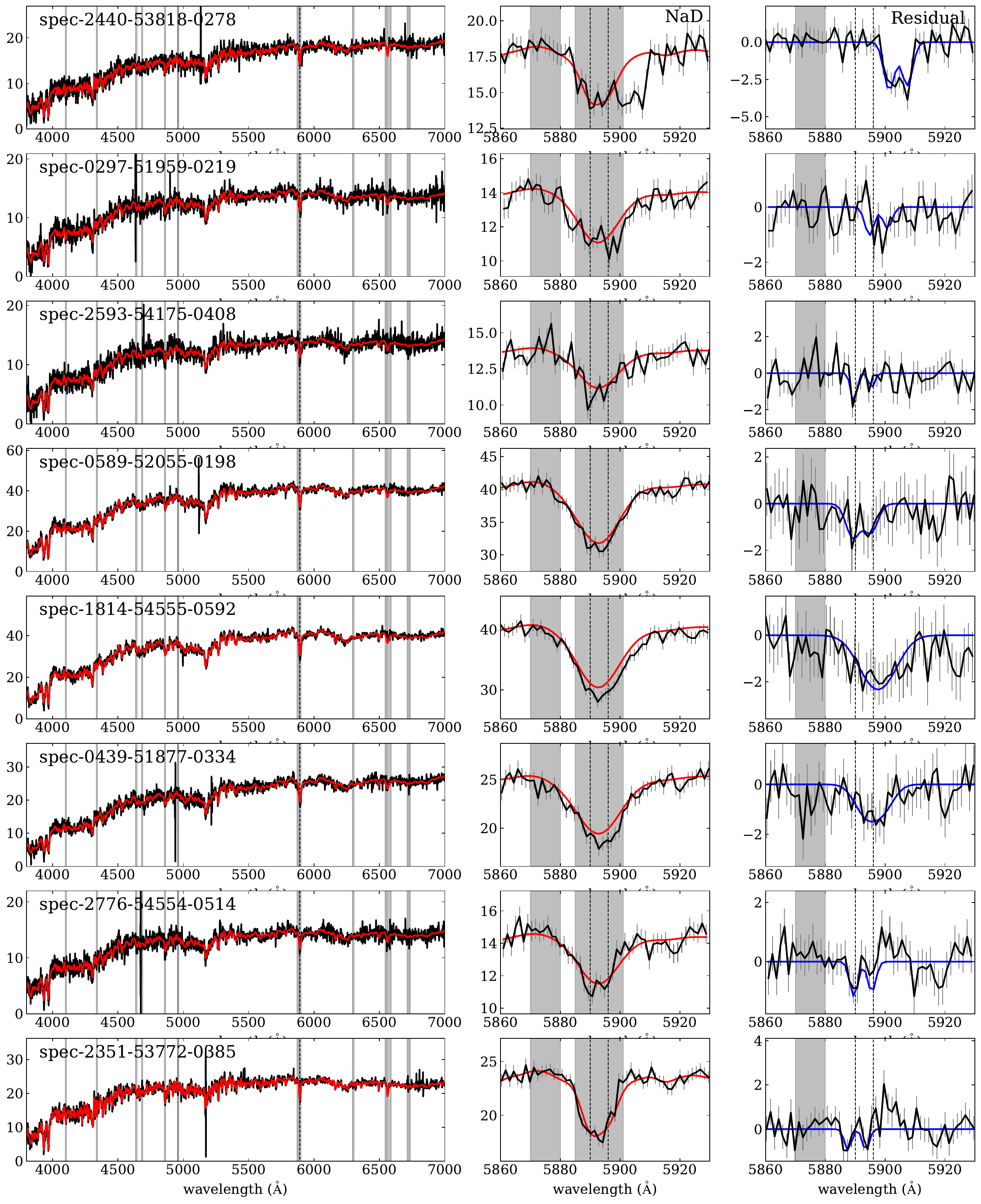}
    \caption{Same as Figure~\ref{Fig:PSB_sample}, but for our quiescent sample.}
    \label{Fig:QSC_sample}
\end{figure*}

In this section, we describe how we fit the $\nad$ doublet absorption line profiles
to derive flow properties. 
Recently, several works studied $\nad$ absorption properties
using the "stacked" SDSS spectra of large galaxy samples
\citep{chen_absorption-line_2010,cicone_outflows_2016,concas_light_2017,roberts-borsani_prevalence_2018}; these analyses produce high $S/N$ spectra, but preclude exploration of the $\nad$ properties of individual galaxies.
Therefore, in this paper, we focus on the $\nad$ properties of each galaxy
in the three galaxy samples from Section~\ref{sec.samp} to learn how those properties evolve in time.
Our analysis consists of three steps:
1) modelling, fitting, and subtracting the stellar continuum, 
2) measuring the EW of the residual $\nad$ absorption, i.e., due to the neutral ISM, as well as the EW of the total $\nad$ absorption before continuum-subtraction, i.e., the sum of the stellar and neutral ISM contributions, and
3) determining the offset of the residual $\nad$ 
from the redshift of the galaxy. 

\subsection{Stellar Continuum Modeling}\label{sec:stellarfit}
We model the stellar continuum of each galaxy 
spectrum over the wavelength range of 3800-7000$~\angstrom$ 
in the rest-frame, using the penalized pixel fitting \citep[pPXF,][]{cappellari_parametric_2004,cappellari_improving_2017} Python code.
We mask some significant emission and absorption lines, e.g., Balmer lines, $\NIII$ $\lambda4640$, $\heii$ $\lambda4686$, [$\OIII$] $\lambda\lambda4959, 5007$, $\hei$ $\lambda5875$, [$\OI$] $\lambda6300$, [$\NII$] $\lambda\lambda6548, 6584$, and [$\SII$] $\lambda\lambda6717, 6731$, as they may affect the best fits of the stellar continuum models. The $\nad$ lines are also masked because they include both stellar absorption and ISM contributions. A polynomial with degree$=4$ is added to avoid mismatches between galaxy spectra and stellar templates and ensure the accuracy of fitted kinematic parameters. 

We use MILES \citep{vazdekis_evolutionary_2010} simple stellar population (SSP) models as stellar templates, adopting the SSP model spectra with four different metallicities: ${\rm [M/H]}=-0.71$, $-0.40$ (sub-solar), $0.00$ (solar), and $+0.22$ (super-solar). For each galaxy, we convert its stellar mass into a range of metallicities using the mass-metallicity relation from \citet{gallazzi_ages_2005} (see Figure 8 in this paper): 

\begin{enumerate}

    \item $\lgmstar<10.0$:
    ${\rm [M/H]}=$ -0.71, -0.40, 0.00; 
    \item $10.0\leq \lgmstar <10.5$: 
    ${\rm [M/H]}=$ -0.71, -0.40, 0.00, $+0.22$;
    \item $10.5\leq \lgmstar <11.0$: 
    ${\rm [M/H]}=$ -0.40, 0.00, $+0.22$;
    \item $\lgmstar\geq11.0$: 
    ${\rm [M/H]}=$ 0.00, $+0.22$.
    
\end{enumerate} 

To determine the impact of the metallicity choice on our results, we also test the assumption of constant solar metallicity for all galaxies, finding that the following results and conclusions do not change significantly. Therefore, we use the results based on the four different metallicities throughout this paper, 
because this choice is more physically motivated. 

pPXF provides the best-fit redshift and its error from the stellar velocity measurement. We use redshifts from the MPA-JHU catalogs as an initial guess and run iterations to obtain the best-fit results from pPXF, although the two redshifts are similar. If the difference between the redshifts of two adjacent iterations is less than $3~\kms$, we consider the results as converged and stop iterating; otherwise, the iteration continues until the total number of iterations reaches 10. 
Finally, we shift the original and the residual (after subtracting the best-fit stellar continuum model) spectra to the rest-frame wavelength given by the pPXF redshift.

The left columns of Figures~\ref{Fig:PSB_sample}, \ref{Fig:SBG_sample}, and~\ref{Fig:QSC_sample} show examples of galaxy spectra and the best-fit stellar continuum models (red lines) for post-starburst, starburst, and quiescent galaxies, respectively. The region immediately around $\nad$ and the best-fit model are shown in the middle columns. The masked wavelength regions are marked as grey areas. Given that pPXF does not provide flux errors for the best-fit stellar models, we assume that the residual flux error is the same as that of the original flux. Systematic errors arising from the assumed stellar population models may thus affect our estimations of the ISM properties. 

\subsection{EW of Excess Na D due to Neutral ISM}\label{sec:EW_analysis}
The total $\nad$ absorption in the original galaxy spectrum arises from a combination of stellar and neutral ISM components. After stellar continuum subtraction, the residual $\nad$ line indicates the amount of absorption due to the neutral ISM.
To quantify the amount of absorption, we measure the total and residual EWs, $\ewtot$ and $\ewexc$, respectively. In this work, a positive EW indicates absorption. To avoid contamination from the $\hei$ $\lambda5875$ emission line frequently seen in star-forming galaxy spectra, we mask 5$~\angstrom$ around 5875$~\angstrom$. 
The EW measurement is performed over the wavelength range 5880-5910$~\angstrom$, and the continuum level is defined as the average of the fluxes at 5850-5870$~\angstrom$ and 5920-5940$~\angstrom$ in the best-fit stellar continuum model. 
We perform a Monte Carlo simulation, generating 1000 mock spectra based on the observed flux errors, to derive the measurement errors of $\ewtot$ and $\ewexc$.

\subsection{Velocity Offset of Na D Excess}\label{sec:vel_analysis}

The next step is to fit the profile of the $\nad$ excess and determine its velocity offset ($\Delta V$) from the galaxy redshift determined by pPXF. Examples of the $\nad$ excess fitting for three galaxy sample groups are shown in the right panels of Figure~\ref{Fig:PSB_sample}, ~\ref{Fig:SBG_sample}, and ~\ref{Fig:QSC_sample}.
We apply a double Gaussian profile to fit the $\nad$ residual spectra. Several previous studies \citep[e.g.,][]{rupke_outflows_2005,chen_absorption-line_2010,lehnert_na_2011,sarzi_cold-gas_2016} used a profile derived by a physical model to fit more physical parameters (the optical depth and the covering factor) and multiple components to distinguish different kinematic parts. Our fitting method is more straightforward to obtain the mean velocities of bulk flows, where a positive $\Delta V$ is an inflow and negative $\Delta V$ is an outflow.   

The double Gaussian profiles that we use for fitting the $\nad$ excess are:
\begin{equation}
    y=p_1~{\rm exp}\left(-\frac{(x-\mu)^2}{2\sigma^2}   \right) +
    p_2~{\rm exp}\left(-\frac{(x-\mu-6)^2}{2\sigma^2}   \right).
\label{eq2}
\end{equation}
Here we assume that the widths of the two
$\nad$ lines are the same ($\sigma$) and that their height ratio ($p_2/p_1$) ranges from 1 to 2, corresponding to optically-thick and thin scenarios, respectively. We also restrict the line widths to be no smaller than the 
instrumental resolution of the SDSS spectra (70 $\kms$). In some cases, the $\nad$ lines are resolved (e.g., the first row of Figure~\ref{Fig:PSB_sample}), while some are blended (e.g., the fifth row of Figure~\ref{Fig:PSB_sample}). The wavelength shift of the $\nad$ excess $\mu$ is varied over the range [$-20$, $20$]$ + 5890~\angstrom$, which
corresponds to $\Delta V \sim \pm$ 1000 $\kms$.  We fit the double Gaussian profiles to the $\nad$ excess at 5850-5940$~\angstrom$ using the {\tt\string PYTHON PYCMPFIT} package\footnote{\url{https://github.com/cosmonaut/pycmpfit}}. 

Depending on the geometry, it is possible to see $\nad$ emission when absorbed photons are re-emitted along the line of sight. This scattering process can produce P-Cygni profiles of $\nad$ (see the profiles of $\MgII$ $\lambda\lambda2796, 2803$ simulated by the cold gas wind models in \citealt{prochaska_simple_2011}). We find that some galaxies show a clear P-Cygni profile in their residual spectra (the two bottom rows of Figures ~\ref{Fig:PSB_sample}, ~\ref{Fig:SBG_sample} and ~\ref{Fig:QSC_sample}). Given that blueshifted absorption is accompanied by redshifted emission, the overall profile should be fit by two components.  

To identify whether the $\nad$ has a pure absorption (or emission) profile or 
a P-Cygni profile, we use two individual double Gaussian profiles to fit the absorption and emission components separately. Then, we classify the profiles as follows:

\begin{enumerate}
    \item Pure Absorption: the peak of only the absorption component has signal-to-noise ratio $\mathrm{S/N}_\mathrm{ab}$ $\geq3$;
    \item Pure Emission: the peak of only the emission component has $\mathrm{S/N}_\mathrm{em}$ $\geq3$;
    \item P-Cygni: both $\mathrm{S/N}_\mathrm{ab}$ and $\mathrm{S/N}_\mathrm{em}$ $\geq3$;
    \item Nothing: both $\mathrm{S/N}_\mathrm{ab}$ and $\mathrm{S/N}_\mathrm{em}$ $<3$.
\end{enumerate}
\begin{table}
	\centering
	\caption{Numbers of galaxies in the starburst, post-starburst, and quiescent samples with different $\nad$ properties and kinematics.}
	\label{Tab:sample_size}
	\begin{threeparttable}
    	\begin{tabular}{cccc} 
    		\hline
    		Classification &  Starburst & Post-starburst & Quiescent\\
    		\hline
    		Total & 3306 & 516 & 72251\\
    		\hline
    		Pure Absorption\tnote{1} & 580 & 189 &  9698  \\
            Pure Emission\tnote{2} & 100 &  95 &  1002  \\
            P-Cygni\tnote{3} & 9 & 11 & 13  \\
            Nothing\tnote{4} &  2617 & 221 & 61538  \\
    		\hline
            Flow-detectable\tnote{5}  & 370  & 163 &  4598 \\
            \hline
            Outflow\tnote{6}  & 280& 105 & 1404 \\ 
            Inflow\tnote{7}   & 90 & 58 & 3194 \\
            \hline
    	\end{tabular}
	\begin{tablenotes}
    \item[1] Galaxies with only $\mathrm{S/N_{ab} > 3}$ $\nad$ excess in residual spectrum.
    \item[2] Galaxies with only $\mathrm{S/N_{em} > 3}$ $\nad$ excess in residual spectrum.
    \item[3] Galaxies with both $\nad$ absorption and emission in residual spectrum: $\mathrm{S/N_{ab} > 3}$ and $\mathrm{S/N_{em} > 3}$.
    \item[4] Galaxies with neither $\nad$ absorption nor emission in residual spectrum: $\mathrm{S/N_{ab} \leq 3}$ and $\mathrm{S/N_{em} \leq 3}$.
    \item[5] ``Pure Absorption" with significant $\nad$ excess ($\ewexc>3\sigma(\ewexc)$) or ``P-Cygni" galaxies for which residual $\nad$ profile is well-fit.
    \item[6] Flow-detectable galaxies with negative $\Delta V$, including some consistent with zero velocity.
    \item[7] Flow-detectable galaxies with positive $\Delta V$, including some consistent with zero velocity.
    \end{tablenotes}
	\end{threeparttable}
\end{table}
The numbers of galaxies identified as “Pure Absorption,” “Pure Emission,” “P-Cygni,” or “Nothing” of the three galaxy samples are shown in Table~\ref{Tab:sample_size}. From here on, we exclude galaxies 
labelled as “Nothing” or “Pure Emission” to focus on deriving the kinematics properties of the neutral gas from the $\nad$ absorption feature.

To identify the subset of galaxies with reliable fits to the excess $\nad$ profile and the bulk velocity, we first require that, for ``Pure Absorption'' cases, the
$\nad$ excess be significant ($\ewexc - 3\sigma(\ewexc) > 0$). We also include all galaxies with “P-Cygni” profiles. We then employ Monte Carlo simulations again, sampling the error distribution of the residual $\nad$ profile for each galaxy, and fitting each mock residual spectrum. For “P-Cygni” cases, we use two double Gaussian profiles to fit the emission and absorption together. If more than half are well-fit, i.e.,
the error for the $\Delta V$ is determined and the parameter values are not at the boundaries, we consider the flow detectable in that galaxy. 
As mentioned in Section~\ref{sec:stellarfit}, we assign errors to the residual spectra from the original flux uncertainties and ignore any arising from the stellar population model. This choice does not affect our conclusions;
even if we relax the restriction from $3\sigma$ to $1\sigma$, we obtain the same main results.
In Table~\ref{Tab:sample_size}, we summarize the number of these ``flow-detectable galaxies'' in the post-starburst, starburst, and quiescent samples.

Table~\ref{Tab:sample_size} shows that the fraction of flow-detectable galaxies is higher for the post-starburst sample (31\%) than for the starburst (11\%) and quiescent (6\%) samples. The lower flow-detectable rate is driven in part by higher fractions of low $S/N$ spectra in the starburst and quiescent samples, making a higher fraction of the residual $\nad$ line profiles difficult to fit.\footnote{ 
Our post-starburst sample comes from F18, who selected their post-starburst sample by requiring the $S/N$ of the SDSS spectra to be $>$10, as we do here. For better post-burst age dating, however, F18 further required that their post-starbursts had good UV detections from \textit{GALEX}, which pushed the mean $S/N$ higher. Therefore, even though all three of our galaxy samples, starburst, post-starburst, and quiescent, are selected with $S/N$ $>$ 10, the prior selection of the post-starburst galaxies by F18 leads to a higher $S/N$ on average for our post-starbursts.
Even though the $S/N$ distributions of three samples are slightly different, our SFR and stellar mass selection criteria ensure that the three samples form a feasible sequence in time, tracing the SFR declining sequence (Figure~\ref{Fig:SFR_sequence}) and the mass growth sequence (see \S\ref{sec:ComparSamp}). In other words, the starburst sample is consistent with being the statistical progenitors of the post-starburst sample, and the quiescent sample is consistent with being the end-products of the post-starburst sample.}
The flow-detectable fraction also depends on $M_*$, which we discuss further in Section~\ref{sec.ISM&flow_detect_result}.

For the flow-detectable galaxies, we measure $\Delta V$ and its measurement uncertainty based on the Monte Carlo simulation results of the $\nad$ excess profile fitting. For the “P-Cygni” cases, $\Delta V$ is measured from the wavelength shift of the absorption component relative to the galaxy's systemic velocity. The final uncertainty in $\Delta V$ is propagated from the dispersion of the line shift measurements from the Monte Carlo simulations and the error in the pPXF redshift.

\begin{figure*}
	\includegraphics[width=0.95\textwidth]{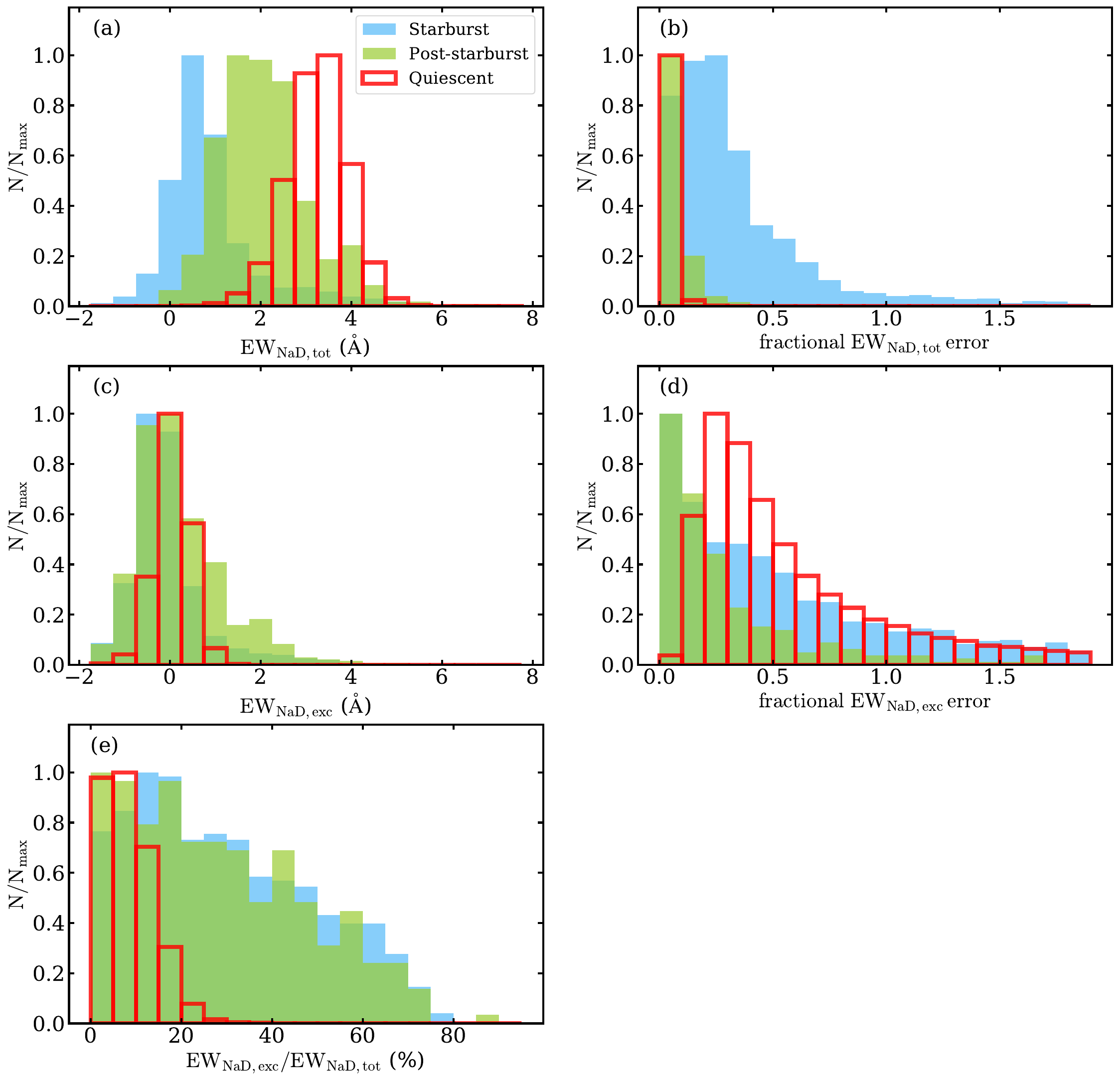}
    \caption{Normalized histograms of (a) the EWs of $\nad$ absorption $\ewtot$, (b) fractional errors on the $\ewtot$, (c) the EWs of $\nad$ excess $\ewexc$, (d) fractional errors on the $\ewexc$, and (e) the ratios $\ewexc$/$\ewtot$ for the three galaxy samples. Blue, green, and red histograms represent the results of the starburst, post-starburst, and quiescent galaxy samples, respectively. Positive EW indicates absorption and vice versa. In panel (e), we plot the ratios only for galaxies with both positive $\ewtot$ and $\ewexc$. The $\ewexc$/$\ewtot$ ratio is relatively small for the quiescent sample. The explanation is suggested by the fact that the median $\ewtot$ increases from starburst to quiescent (see panel (a)), but the peaks of the $\ewexc$ distributions for all three galaxy samples are similar, i.e., around 0$~\angstrom$ (panel (c)). Therefore, the low ratio of the quiescent sample arises from a low fraction of neutral gas to stars.}
    \label{Fig:hist_EW}
\end{figure*}

\section{Results}\label{sec.results} 

\subsection{ISM Contribution to Na D: 
$\ewexc/\ewtot$}\label{sec.ew_result} 

The three different galaxy samples in our evolutionary sequence of declining star formation have 
different $\ewtot$ distributions (panel (a) of Figure~\ref{Fig:hist_EW}). 
The median values of $\ewtot$ from starburst to post-starburst to quiescent galaxies increase from 0.68$~\angstrom$ to 1.99$~\angstrom$ to 3.27$~\angstrom$ as the dominant stellar population changes from early-type stars to late-type stars and results in stronger $\nad$ stellar absorption.
The median fractional error in $\ewtot$ 
is 23\% for the starburst sample, 6\% for 
the post-starburst sample, and 4\% for the quiescent sample (panel (b) of Figure~\ref{Fig:hist_EW}).  

Panel (c) of Figure~\ref{Fig:hist_EW} shows the distributions of the $\ewexc$ for the three galaxy samples. All three histograms have peaks at around 0$~\angstrom$. The median $\ewexc$ is -0.2$~\angstrom$, 
0.01$~\angstrom$, and 0.07$~\angstrom$ for the starburst, post-starburst, and quiescent samples, respectively. The histograms for the starburst and the post-starburst samples have tails towards positive $\ewexc$, but not for the quiescent sample. The median fractional error on $\ewexc$ is 47\%, 25\%, and 56\% for the starburst, post-starburst, and quiescent samples, respectively, as shown in panel (d) of Figure~\ref{Fig:hist_EW}. 

Panel (e) of Figure~\ref{Fig:hist_EW} shows the distributions of the ratio $\ewexc/\ewtot$ for the three galaxy samples. The $\ewexc/\ewtot$ ratio traces the ISM contribution to the total $\nad$ absorption line. In this panel, we plot only the galaxies with positive $\ewtot$ and $\ewexc$: 1,139 starbursts, 262 post-starbursts, and 41,422 quiescent galaxies. Most (97\%) of the quiescent galaxies have $\ewexc/\ewtot$ smaller than 20\%, compared 39\% for starburst galaxies and 41\% for post-starburst galaxies. The median 
$\ewexc/\ewtot$ is 8\% for the quiescent galaxies, 
27\% for the starbursts, and 25\% for the post-starbursts.
Given the fact that the median $\ewtot$ increases from starburst to quiescent (see panel (a)), but the peaks of the $\ewexc$ distributions for all three galaxy samples are similar (panel (c)), the low 
$\ewexc/\ewtot$ ratios typical of the quiescents arise from a low fraction of neutral gas to stars.

\subsection{Neutral Gas Detection versus Stellar Mass}\label{sec.ISM&flow_detect_result} 

\begin{figure*}
    \includegraphics[width=0.95\textwidth]{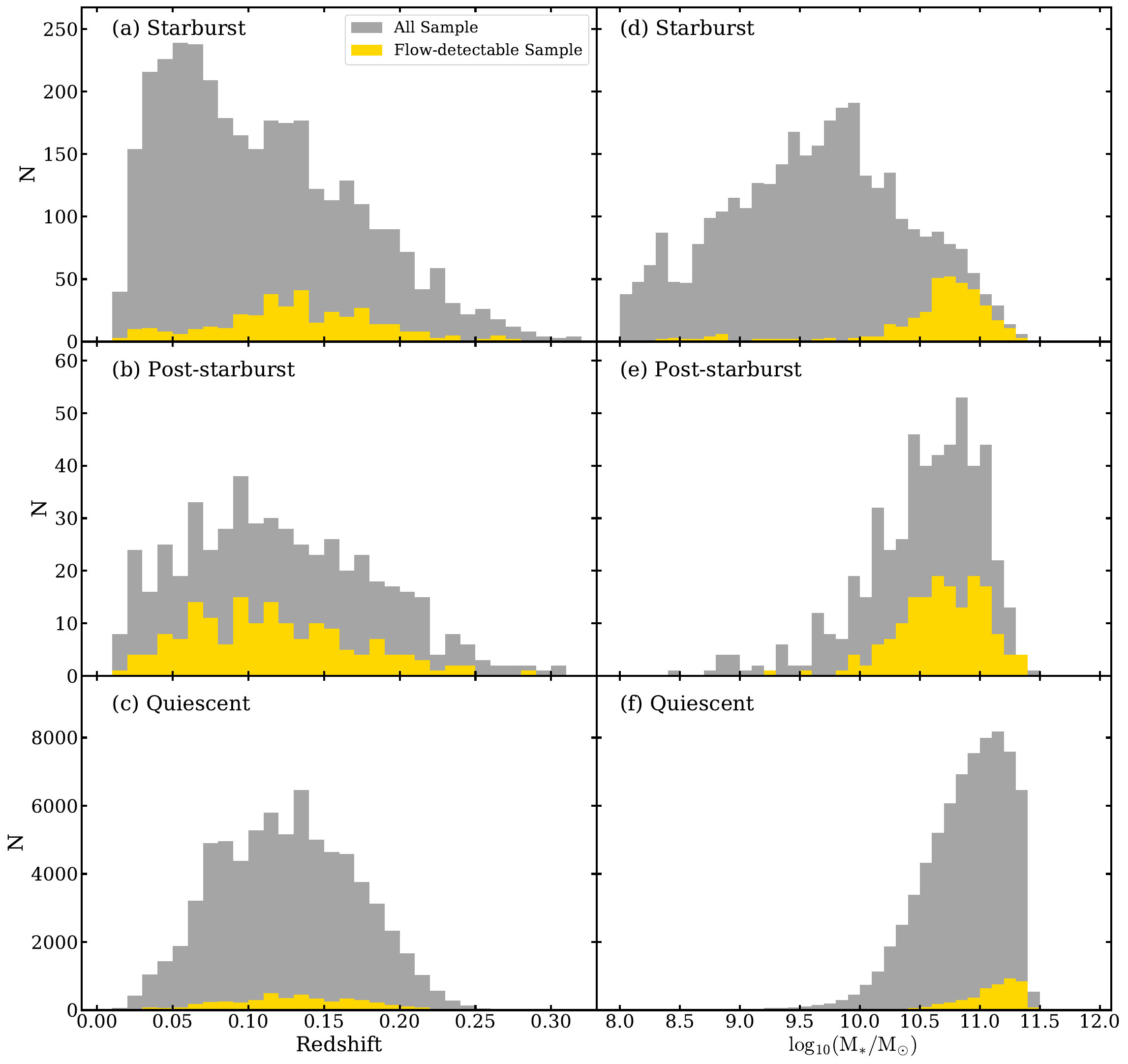}
  \caption{Histograms of the redshift (left) and stellar mass (right) for the three galaxy samples. Gray: All galaxies; 
  Yellow: Flow-detectable sample, i.e, with a significant “Pure Absorption” ($\ewexc>3\sigma(\ewexc)$) or a “P-Cygni” $\nad$ excess, and a good velocity measurement. The flow-detectable galaxies in the three galaxy samples are distributed over the entire redshift range, but they lie mostly at higher stellar masses ($\lgmstar>10$). Even controlling for the effects of $S/N$, the rates are higher in higher mass galaxies.}
    \label{Fig:hist_Mstar_z}
\end{figure*}

We test the dependencies of the flow-detectable sample in Table~\ref{Tab:sample_size} on $M_*$. The right column of Figure~\ref{Fig:hist_Mstar_z} shows that, for all three types of galaxies, the rate of flow-detectables increases with stellar mass, with most galaxies having $\lgmstar>10$. 
Even controlling for the effects of $S/N$ (see Section~\ref{sec.wind_prop_trend}), the rates are higher in higher-mass galaxies.

This result is consistent with the findings of \citet{roberts-borsani_prevalence_2018}, who found neither inflows nor outflows in their low-mass galaxies. 
This trend arises because $\nad$ is very fragile---its low ionization potential (5.14 eV) requires it to be shielded from hydrogen-ionizing photons by a large amount of dust. In more massive galaxies, we would expect to detect the $\nad$ ISM feature, because the metal-richness and dust lead to better shielding of $\nad$. 
\citet{roberts-borsani_prevalence_2018}
also found that the $\nad$ residual profiles of their low-mass non-AGN host galaxies ($\lgmstar\leq10$) are typically in emission, but we exclude
such ``Pure Emission" galaxies from our flow-detectable sample (see Section~\ref{sec:vel_analysis}). 

The left column of Figure~\ref{Fig:hist_Mstar_z} shows the distributions of redshift for the three galaxy samples; the flow-detectable rate is generally independent of redshift.

\subsection{Significant Inflows versus Outflows}\label{sec.vel_result} 

\begin{figure}
    \includegraphics[width=\columnwidth]{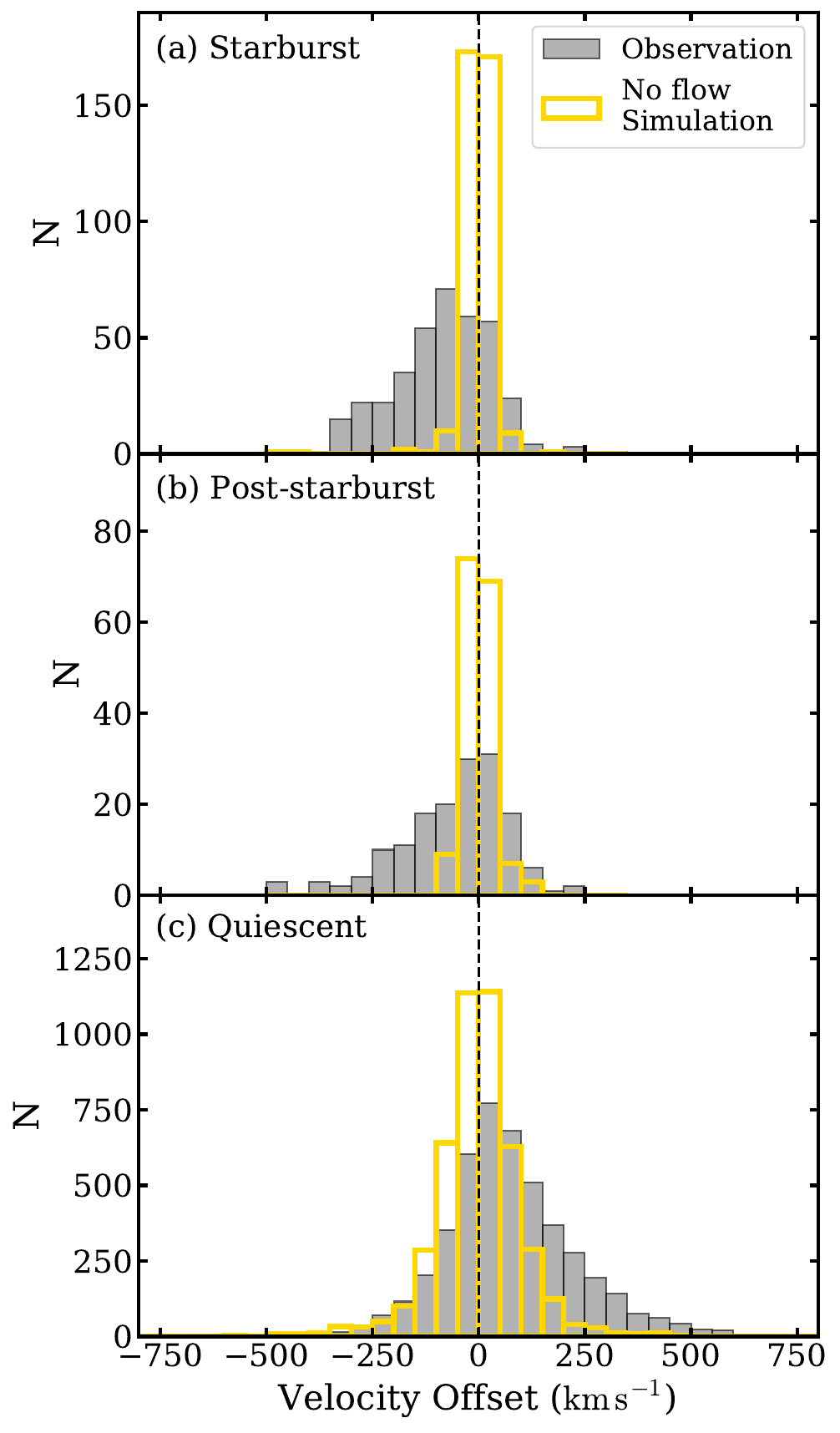}
\caption{Histograms of velocity offset $\Delta V$ for the (a) starburst, (b) post-starburst, and (c) quiescent samples. The observed $\Delta V$ distributions are displayed by filled gray histograms. The open yellow histograms are the simulated distributions expected for galaxies without flows, broadened by the 1$\sigma$ measurement errors. The mean velocity offsets and their errors for the three galaxy samples from top to bottom are $-84.6\pm5.9~\kms$, $-71.4\pm11.5~\kms$, and $76.6\pm2.3~\kms$. For the starburst and post-starburst samples, the observed distributions show significant negative velocity tails compared to the static simulations, indicating the presence of outflows. The quiescent sample has a significant positive velocity tail, revealing the prevalence of inflows. }
\label{Fig:zero_vel_simu}
\end{figure}

Hereafter we focus only on the 163 post-starburst 
galaxies, 370 starburst galaxies, and 4598 quiescent galaxies that are flow-detectable.
Assuming a possible evolutionary sequence from starburst to post-starburst to quiescent, Figure~\ref{Fig:zero_vel_simu} shows the histograms of $\Delta V$ for the three samples defining this sequence (grey histograms). The histograms for the starburst and post-starburst samples are significantly skewed with a tail at high negative velocity offsets. The mean velocity offsets are $-84.6\pm5.9~\kms$ and $-71.4\pm11.5~\kms$ for the starburst and post-starburst samples, respectively. On the other hand, the velocity offset distribution of the quiescent sample
is skewed to the positive side with the mean velocity offset of $76.6\pm2.3~\kms$. The uncertainty of the mean velocity offset is calculated from the overall dispersion of velocity offset in each galaxy sample.

To test whether these differences are statistically significant, we compare the distribution of observed $\Delta V$ with the 
distribution expected for galaxies with zero offsets, broadened by the $1\sigma$ measurement errors (yellow histograms in Figure~\ref{Fig:zero_vel_simu}). For the starburst and post-starburst samples, the $\Delta V$ distributions show negative tails that exceed the distributions expected in the absence of outflows. This difference is confirmed by a Kolmogorov-Smirnov (K-S) test.
On the other hand, the distribution of $\Delta V$ for the quiescent sample shows a significant excess of positive velocity offsets, indicating inflows.
To test the effect of underestimating the $\Delta V$ errors, we also broaden the zero flow distribution assuming the $1.5\sigma$ and the $2\sigma$ errors; the excess tails remain and are still significant.

Assuming that the $\nad$ excess represents the average properties from the foreground ISM in the sightline, negative and positive velocities represent ISM
bulk flows in opposite directions: negative velocities indicate outflows, while positive velocities indicate inflows\footnote{We do not use the 1$\sigma$ uncertainty to constrain the outflow/inflow definition, as doing so may lead to different bulk motion strength cuts for different galaxy samples, which might bias our subsequent analysis.}. Therefore, the excess negative velocity tails of the starburst and post-starburst samples indicate significant outflow populations, while the positive tail of the quiescent sample suggests a significant inflow population. The numbers of outflows and inflows are shown in Table~\ref{Tab:sample_size}.

\subsection{Trends with Evolutionary Sequence}\label{sec.wind_prop_trend} 
\subsubsection{Velocity offset and outflow fraction from starburst to quiescent}\label{sec.wind_prop_sequence}

\begin{figure*}
\centering
    \includegraphics[width=0.9\textwidth]{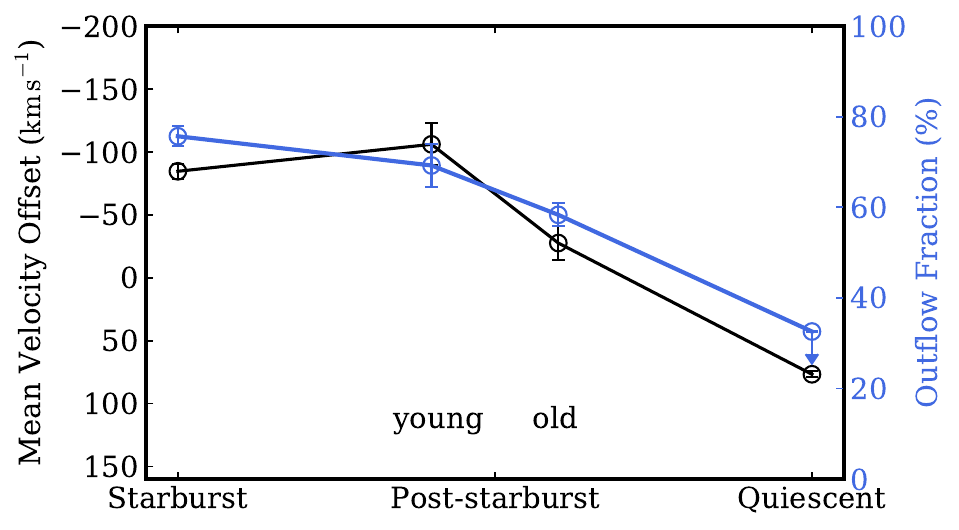}
\caption{Trends of mean velocity offset (black) and outflow fraction (blue) along the evolutionary sequence from starburst to post-starburst to quiescent galaxies. The post-starburst sample is divided into two subsamples: young (post-burst age $< 400\,\mathrm{Myr}$) and old ($\geq 400\,\mathrm{Myr}$). Mean velocity offset changes from negative to positive along the sequence from $-84.6\pm5.9$ to $-106.0\pm16.8$ to $-27.6\pm13.3$ to $76.6\pm2.3\,\kms$. 
There is a similar trend from old to young post-starbursts that is significant at $>2\sigma$. The change in outflow fraction is qualitatively similar, from $76\pm2\%$ to $69\pm5\%$ to $58\pm3\%$ to $33\%$, where the latter is a 3$\sigma$ upper-limit. These trends suggest that outflows diminish as galaxies age and star formation rate decreases.
}
\label{Fig:vel_outflow_fraction_sequence_tot}
\end{figure*}

\begin{figure*}
\centering
    \includegraphics[width=\textwidth]{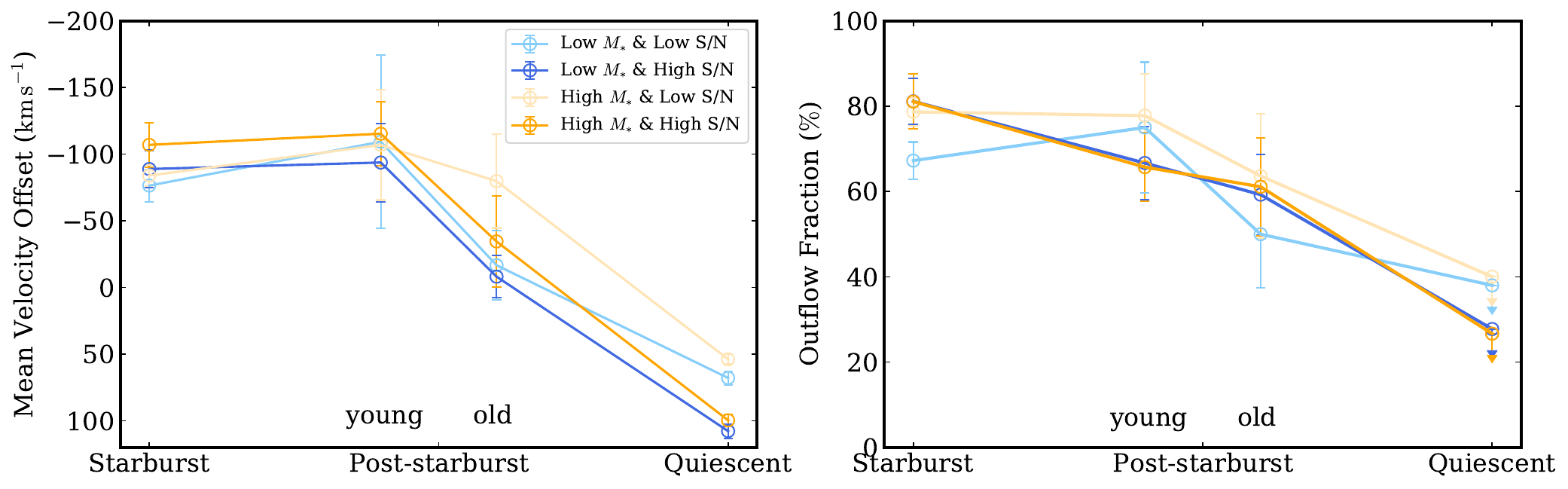}
\caption{Trends of mean velocity offset (left) and outflow fraction (right) along the evolutionary sequence in four
stellar mass and $S/N$ bins. The light-blue and dark-blue curves are for low stellar mass galaxies with low and high $S/N$ spectra, respectively. The light-orange and dark-orange curves are for high stellar mass galaxies with low and high $S/N$ spectra, respectively. The outflow fractions of the quiescents are 3$\sigma$ upper limits. The trends of outflow fraction and mean velocity offset in these four bins are similar to those in Figure~\ref{Fig:vel_outflow_fraction_sequence_tot}, with high stellar mass galaxies shifted more toward negative velocity offsets.
}
\label{Fig:SNR_Mstar_group_seq}
\end{figure*}

In this section, we test how gas flow properties (i.e., $\Delta V$ and the fraction of galaxies with outflows) evolve along the assumed evolutionary sequence of starburst to post-starburst to quiescent. For the post-starburst sample, we further divide it into two groups: young post-starburst galaxies (post-burst age $< 400\,\mathrm{Myr}$) and old post-starburst galaxies (post-burst age $\geq 400\,\mathrm{Myr}$).

As mentioned in Section~\ref{sec.vel_result} and shown in Figure~\ref{Fig:vel_outflow_fraction_sequence_tot}, the mean velocity offset changes from negative to positive along the evolutionary sequence, suggesting a transition from outflows to inflows with diminishing star formation rate and age.
The same trend is seen with post-burst age within the post-starburst sample.
The value of the old post-starburst sample ($-27.6\pm13.3~\kms$) is lower than that of the young sample. 
While we average positive and negative velocities in computing the means, the observed trend is driven by outflows declining along the evolutionary sequence, as shown by a comparison of the panels in Figure~\ref{Fig:zero_vel_simu}. (For the quiescent sample, there is also a contribution from real inflows.)
Hereafter we use the terms ``mean velocity offset'' and ``mean outflow velocity'' to denote overall flow velocity (in any direction or static) and outflow velocity (negative only), respectively.

To test the effects of the stellar mass and spectral $S/N$ on the mean velocity offset trend in Figure~\ref{Fig:vel_outflow_fraction_sequence_tot}, we further divide our samples into four bins: 1) low $M_{*}$ and low $S/N$; 2) low $M_{*}$ and high $S/N$; 3) high $M_{*}$ and low $S/N$; 4) high $M_{*}$ and high $S/N$. The stellar mass subgroups are separated using median values, which are $\lgmstar=10.7$ for the starburst, young post-starburst, and old post-starburst samples, and $\lgmstar=11.0$ for the quiescent sample. The low and high $S/N$ subgroups are separated using the median value $S/N = 20$. 

The trends for the four groups are shown in the left panel of Figure~\ref{Fig:SNR_Mstar_group_seq}. In all of them, the mean velocity offset changes from negative
to positive along the sequence. For the starburst and post-starburst samples, the behavior of the low and high $S/N$ groups is consistent. For the quiescent sample, the low $S/N$ groups lie closer to zero mean velocity offset; this might be due to the general lack of an ISM in quiescent galaxies, which makes it difficult to perform $\nad$ excess fitting and obtain accurate velocity measurements, especially at low $S/N$. 
Or, the less positive mean $\Delta V$ could arise physically, from diminished inflows or more outflows.

The groups divided by stellar mass show similar trends to the overall trend in Figure~\ref{Fig:SNR_Mstar_group_seq}, except that the high $M_*$ groups shift upward at fixed $S/N$. This shift is larger for the high $S/N$ group, suggesting that it is physical. We recover a similar trend for individual post-starburst galaxies in  Section~\ref{sec.vwind_age}.

The outflow fraction is defined as the number of galaxies with outflows as a percentage of the number of flow-detectable galaxies. The outflow fractions for the starburst, post-starburst, and quiescent samples are $76\pm2\%$, $64\pm6\%$, and $33\%$ (3$\sigma$ upper limit), respectively. Considering that outflows in the quiescent galaxies are hard to distinguish from static flows (panel c of Figure~\ref{Fig:zero_vel_simu}), the outflow fraction for the quiescent sample is an upper limit.

\begin{figure*}
    \includegraphics[width=0.9\textwidth]{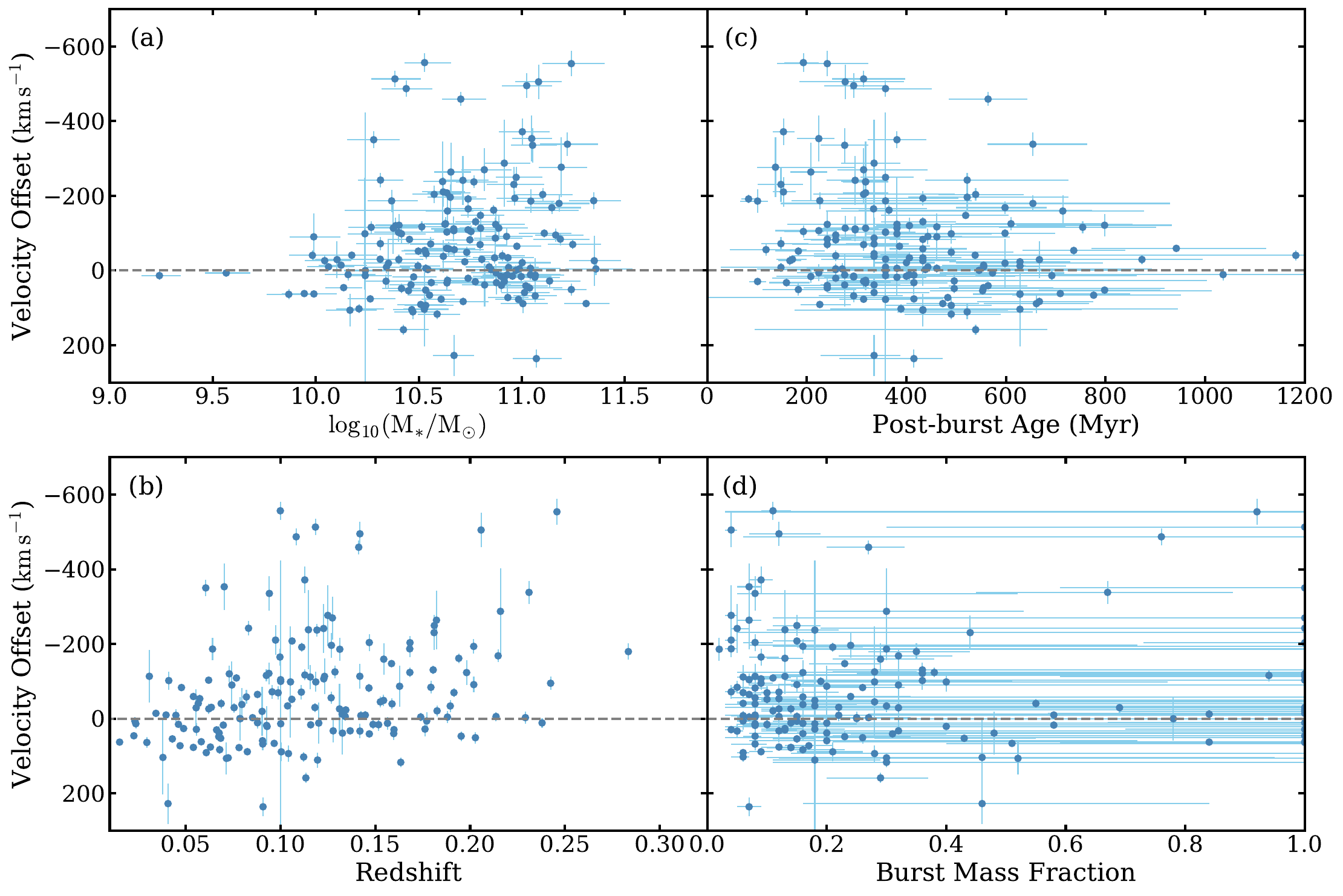}
    \caption{Velocity offset $\Delta V$ of the post-starburst sample as a function of (a) stellar mass, (b) redshift, (c) post-burst age, and (d) burst mass fraction. The Spearman correlation coefficients and their significance values for each panel are (a) $\rho=-0.15,~1.87\sigma$, (b) $\rho=-0.33,~4.25\sigma$, (c) $\rho=0.22,~2.80\sigma$, and (d) $\rho=0.03,~0.34\sigma$, respectively, indicating that the mean $\Delta V$ becomes marginally more negative (toward faster outflows) with increasing stellar mass, more negative with increasing redshift, less negative with increasing post-burst age, and has no dependence on burst mass fraction.}
    \label{Fig:original_vel_trend}
\end{figure*}

Figure~\ref{Fig:vel_outflow_fraction_sequence_tot} displays the overall trend of outflow fraction along the sequence, with the post-starburst sample divided into the young and old subsamples as before. The outflow fraction of the young group is $70\pm5\%$ and that of the old one is $58\pm3\%$. 
In a manner similar to the change in the mean velocity offset,
the outflow fraction decreases along the sequence. As before, we check the stellar mass and $S/N$ dependencies, which are shown in the right panel of Figure~\ref{Fig:SNR_Mstar_group_seq}. For all the stellar mass and $S/N$ groups, the outflow fraction shows a declining trend similar to that in
Figure~\ref{Fig:vel_outflow_fraction_sequence_tot}. 
The lower $S/N$ groups have higher (upper limit) outflow fractions for the quiescent sample. As discussed earlier, this effect may arise because more fake ``outflows'' are detected in quiescents with low $S/N$.

\begin{figure*}
\centering
    \includegraphics[width=0.9\textwidth]{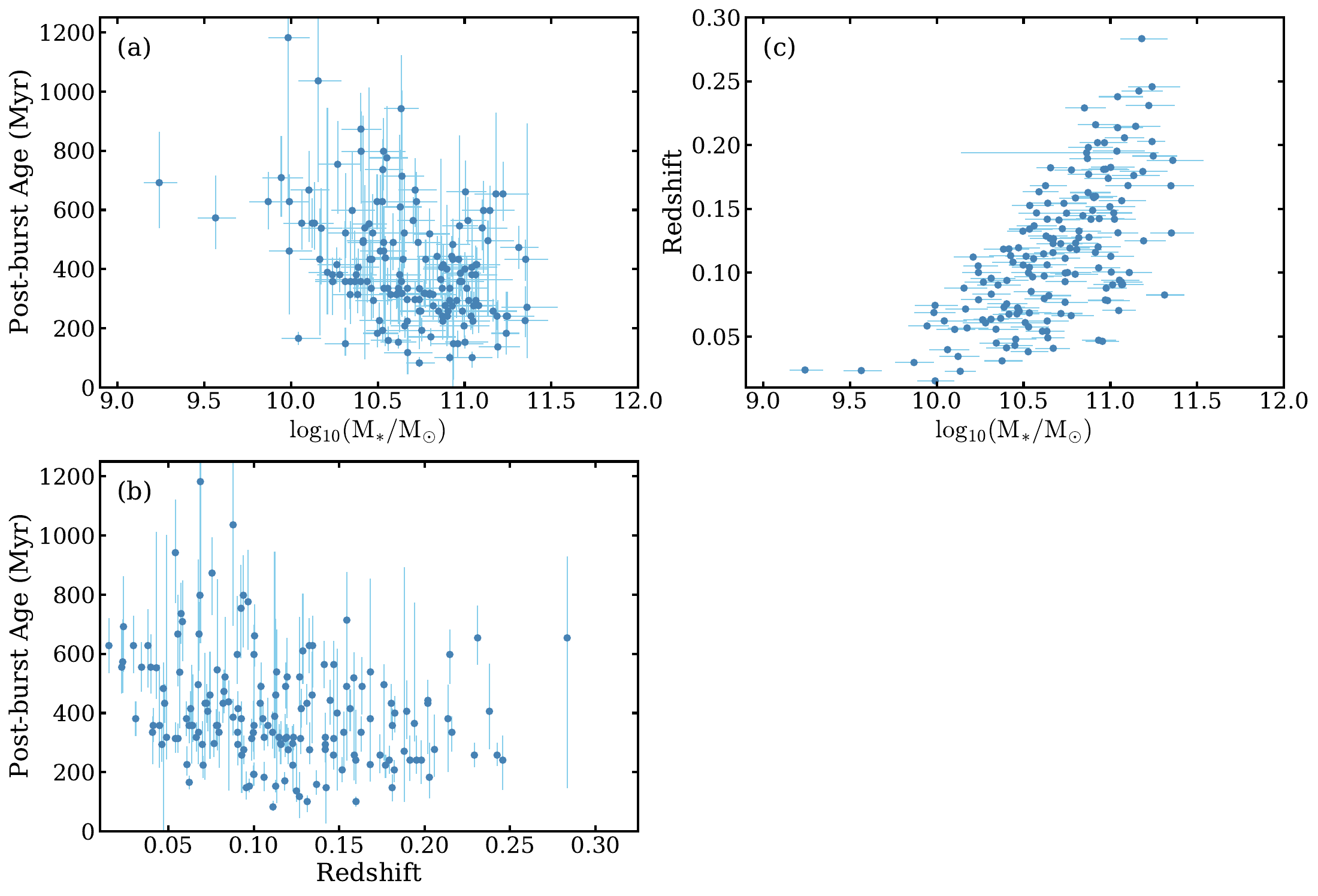}
\caption{(a) Post-burst age vs. stellar mass; (b) Post-burst age vs. redshift; (c) Redshift vs. stellar mass. The Spearman correlation coefficients and their significance values for each panel are (a) $\rho=-0.35,~4.56\sigma$, (b) $\rho=-0.29,~3.78\sigma$, and (c) $\rho=0.64,~9.27\sigma$, respectively. The post-burst age is significantly anti-correlated with both stellar mass and redshift. Moreover, stellar mass and redshift are strongly correlated with each other. Given that stellar mass tracks redshift, we only control for stellar mass when assessing the significance of the velocity offset vs. post-burst age correlation through partial correlation analysis.}
\label{Fig:original_trend_mstar_psbage_z}
\end{figure*}

\subsubsection{Velocity offset vs. age for post-starbursts}\label{sec.vwind_age}

In this section, we focus on the post-starburst phase to consider the correlation between the velocity offset and the post-burst age for individual galaxies. We use the {\tt\string PYTHON Pingouin} package\footnote{\url{https://pingouin-stats.org/build/html/index.html}} to calculate the Spearman partial-correlation coefficient that describes the correlation with the effect of controlling variables removed. The controlling variables are carefully considered to ensure that the correlation between velocity offset and post-burst age is not driven by other factors.

Figure~\ref{Fig:original_vel_trend} shows the $\Delta V$ of the post-starburst sample as a function of stellar mass, post-burst age, redshift, and burst mass fraction. The corresponding Spearman correlation coefficients\footnote{We define a correlation as ``significant'' when the p-value is less than 0.05, i.e., the significance is higher than $2\sigma$ for a normal distribution.} are listed in the caption. The figure indicates that the mean $\Delta V$ becomes less negative with increasing post-burst age ($\rho=0.22,~2.80\sigma$), suggesting that outflows slow in time. The mean $\Delta V$ becomes more negative with increasing redshift ($\rho=-0.33,~4.25\sigma$) and marginally more negative with increasing stellar mass ($\rho=-0.15,~1.87\sigma$). There is no observed dependence of mean $\Delta V$ with burst mass fraction ($\rho=0.03,~0.34\sigma$). 

For the partial-correlation coefficient, only factors correlated with both post-burst age and mean $\Delta V$ need to be considered as controlling variables. Therefore, we do not consider the burst mass fraction because it is $\sim$constant with $\Delta V$. We then test the correlations among stellar mass, redshift, and post-burst age, which are shown in Figure~\ref{Fig:original_trend_mstar_psbage_z}. The post-burst age has significant anti-correlations with stellar mass ($\rho=-0.35,~4.56\sigma$) and redshift ($\rho=-0.29,~3.78\sigma$). 
There is also a correlation between stellar mass and redshift ($\rho=0.64,~9.27\sigma$), which is caused by the magnitude limit of the SDSS sample, where galaxies at higher redshift will have higher average stellar masses.
The strong correlation between stellar mass and redshift suggests that the correlation of post-burst age with stellar mass, and the correlation with redshift, are highly overlapping. Therefore, we only consider stellar mass, which is more likely to be physically correlated with galaxy internal properties, as a controlling variable.

The Spearman partial-correlation coefficient is applied to characterize the correlation between $\Delta V$ and post-burst age after controlling for stellar mass. The final partial-correlation coefficient is 0.18, with $2.29\sigma$ significance. Therefore, there is a significant correlation between $\Delta V$ and post-burst age, even after controlling for stellar mass. This residual correlation implies that outflows diminish during the evolution of the post-starburst phase.

\subsection{Trends with Disk Inclination for Starbursting Galaxies}\label{sec.inclin_result}
\begin{figure*}
\centering
    \includegraphics[width=0.95\textwidth]{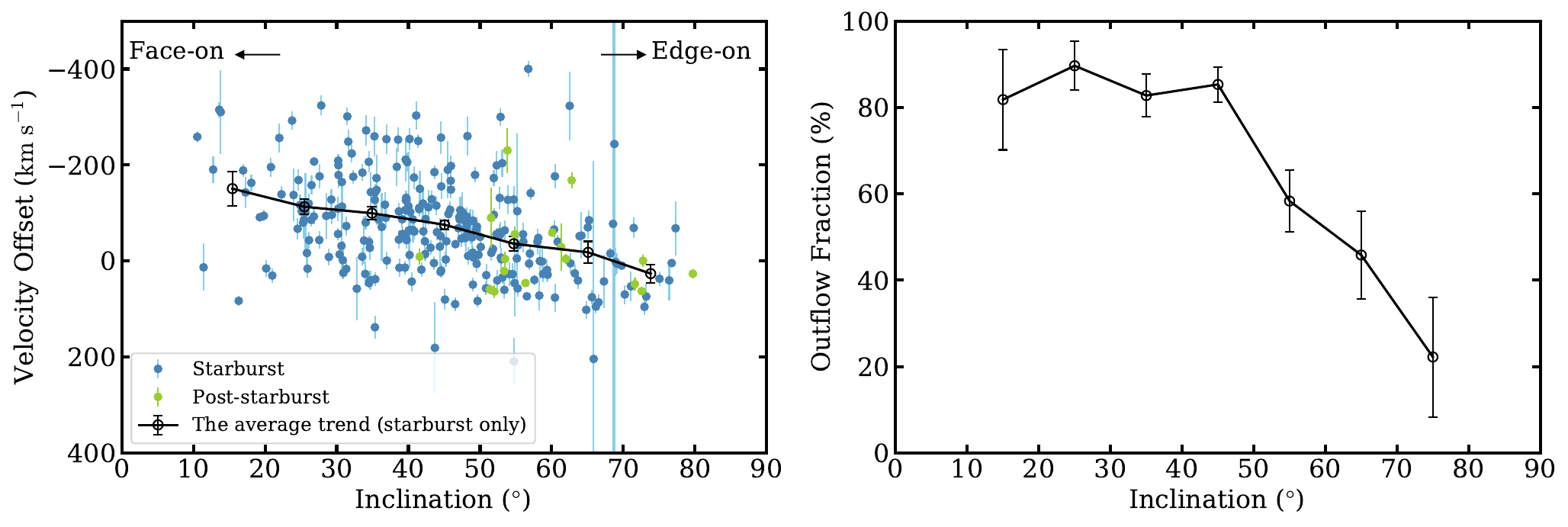}
\caption{Left: Velocity offset vs. inclination. The black line represents the average trend for the starburst sample (blue points); the mean $\Delta V$ becomes less negative with increasing inclination. The declining slope is significantly ($>6\sigma$) different than zero, favoring a star formation-driven wind model in which gas is ejected perpendicularly from the stellar disc plane. 
A similar analysis is challenging for the post-starburst sample (green points), where fewer galaxies have measured inclinations. However, the post-starburst galaxies follow the declining trend of the starburst sample. Right: Outflow fraction vs. inclination for the starburst sample. The outflow fraction, constant at $\sim90\%$ for disc inclinations $i\leq45\degree$, steadily decreases from $\sim90\%$ to $20\%$ for $i>45\degree$.}
\label{Fig:trend_inclin}
\end{figure*}

\begin{figure}
\centering
    \includegraphics[width=0.95\columnwidth]{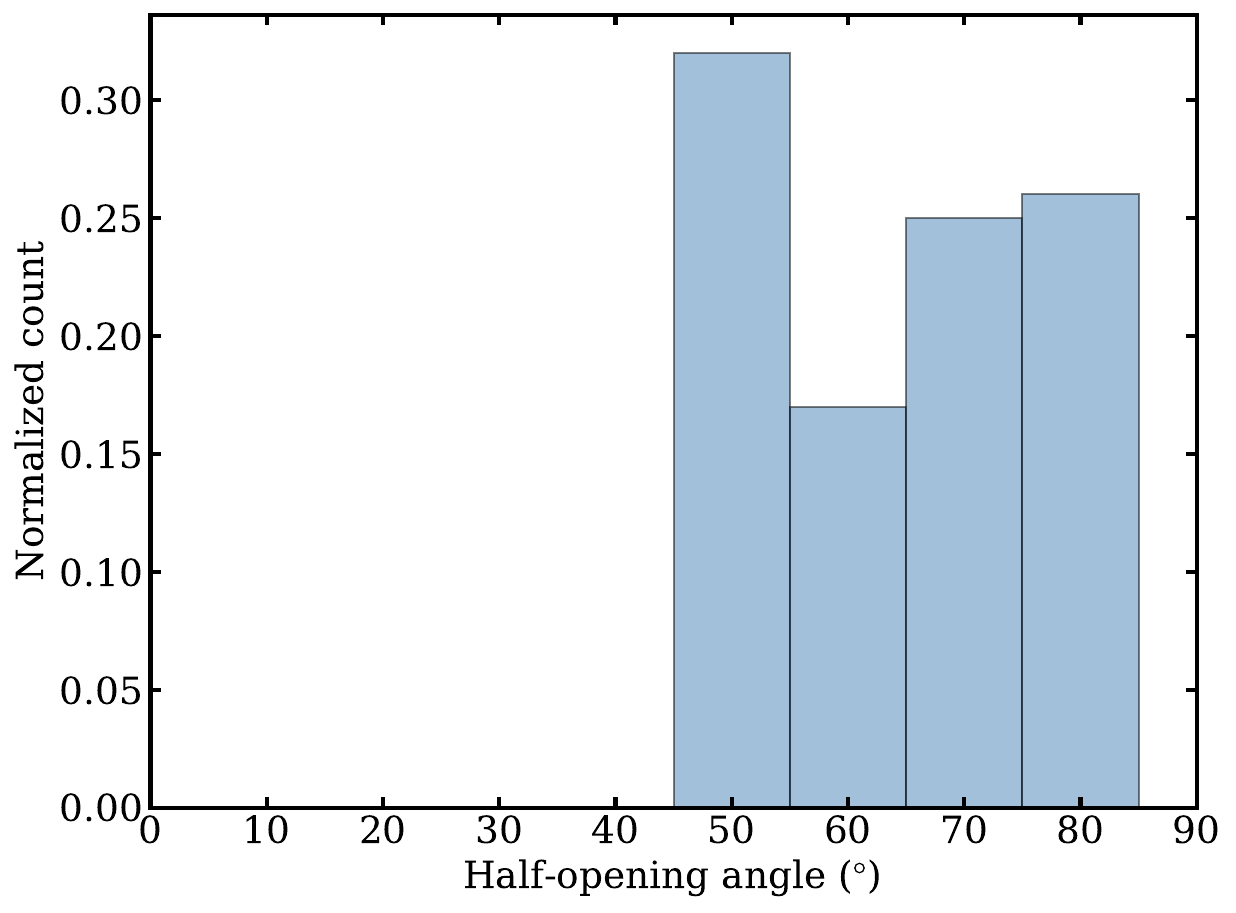}
\caption{Simulated distribution of half-opening angle based on the observed trend of outflow fraction with inclination in Figure~\ref{Fig:trend_inclin}. Our analysis limits the outflow half-opening angle to roughly $>45\degree$.}
\label{Fig:hist_opengingangle}
\end{figure}
In this section, we test the inclination dependency of the mean velocity offset and the outflow fraction to explore the possible mechanism for outflows in starburst galaxies, the only one of our three galaxy samples with a large number of measured disk inclinations.
Galactic outflows usually show a bi-conical structure, i.e., the alignment of outflows is perpendicular to the stellar or AGN accretion disk, and the outflow can only be detected within the opening angle. Therefore, the measured wind velocity along the line of sight depends on the viewing angle. The fastest velocity would be measured when the viewing angle is zero, i.e., aligned with the outflow.
The velocity decreases with the viewing angle due to the projection effect, until the viewing angle exceeds the outflow opening angle and the outflow can no longer be detected.

How the observed velocity offset and outflow fraction vary with galactic disc inclination depends on the origin of the winds (e.g., \citealt{bae_independence_2018,concas_two-faces_2019}). 
These papers find that the velocities of ionized outflows from AGN do not depend on host galaxy inclination, whereas star formation-driven winds are observed to lie principally perpendicular to the disc.
Star formation-driven outflows, which are associated with supernovae and stellar winds, are initially isotropic and then preferentially escape along the galaxy's minor axis, the path of least resistance. In contrast, outflows driven by an AGN jet are perpendicular to the black hole's accretion disc, which can be randomly oriented relative to the host galaxy's disc \citep{veilleux_galactic_2005}. Assuming that an AGN-driven outflow is not dramatically affected by the disc and retains much of its initial orientation and velocity (\citealt{2022AJ....163..134T}; but also see \citealt{2018MNRAS.476.2288H} and \citealt{2019MNRAS.490.3234N}), 
there will be no dependence of outflow velocity on disc inclination for a statistical sample.

The SDSS photometric pipeline \citep{2001ASPC..238..269L} fits each galaxy image with a two-dimensional model of a de Vaucouleurs and an exponential profile. The axial ratios ($b/a$) from the two models are derived by this fitting procedure, but we only use the exponential profile axial ratio below. The pipeline also calculates the best linear combination of the exponential and de Vaucouleurs models as a parameter called $\mathrm{fracDeV}$. Because inclination is only meaningful for disc galaxies, we require $\mathrm{fracDeV} < 0.8$, which corresponds to the disc component being more than 20\% of the total light. In this choice, we follow \citet{2008MNRAS.388.1321P} and \citet{chen_absorption-line_2010}.
The $b/a$ is converted to the inclination using:
\begin{equation}
    \cos ^{2} i=\frac{(b / a)^{2}-q^{2}}{1-q^{2}},
\end{equation}
where $q=0.13$ is the intrinsic axis ratio \citep{giovanelli_extinction_1994}. The typical $b/a$ error is $<0.1$, bringing the error in inclination to about a few degrees. Due to our requirements here that a galaxy has a measured $b/a$ and $\mathrm{fracDeV}$, and then a confirmed disc, we are only able to measure inclinations for 255 starburst and 18 post-starburst galaxies from our samples.

The correlation between $\Delta V$ and inclination for the starburst sample is shown in the left panel of Figure~\ref{Fig:trend_inclin}. The mean velocity offset of the starburst sample 
decreases with inclination and becomes zero when nearly edge-on ($i \sim 80 \degree$). We also do linear regression on this trend using {\tt\string PYTHON LINMIX} package\footnote{\url{https://github.com/jmeyers314/linmix}}, which confirms that the declining slope is significantly($>6\sigma$) different than zero. 

The declining velocity trend with inclination implies that the alignment of outflows detected in the starburst galaxies is more out of the disc, favoring the star formation-driven wind scenario.
It is not surprising that star formation drives winds in our starburst sample---we select for star-forming galaxies in the BPT diagram and against even weak AGN features. \citet{concas_two-faces_2019} also explore the inclination dependence of outflow velocity, but for
galaxies with higher star formation rates than ours: a lower-limit of 12.5 vs. an average of 6.6 $M_{\odot}~\mathrm{yr^{-1}}$ (see Figure~\ref{Fig:SFR_sequence}), respectively. (Some of their galaxies include AGN.) Another difference is that they measure their wind velocities from stacked spectra, instead of for individual galaxies as we do. The slope of their best wind velocity-inclination relation is nevertheless consistent with ours.

For our \emph{post}-starburst galaxy sample, 
there are relatively fewer discs, and those with measured inclinations are concentrated at high values.
As a result, we cannot determine their  $\Delta V$-inclination relation and test for the dominant wind-driving mechanism via wind geometry.  We note, however, that 
the disky post-starbursts in our sample 
do not obviously deviate from the relation defined by our starburst sample.

The outflow fraction trend with inclination for the starburst sample is displayed in the right panel of Figure~\ref{Fig:trend_inclin}. Considering the inclination error, we use an inclination bin size $=10 \degree$. At $i \leq 45 \degree$, the outflow fraction is constant ($\sim$90\%), and outflows are ubiquitous in starburst galaxies. For $i > 45 \degree$, the outflow fraction steadily decreases.
Because randomly distributed outflow alignments would allow us to detect outflows at any inclination with equal probability, the alignment of neutral outflows seen here is more likely to be along the minor axis.

The observed outflow fraction trend with inclination also provides some constraints on the distribution of outflow half-opening angles. First, we assume simplistically that the outflow alignments are exactly perpendicular to the disk. We then assume that the distribution of the half-opening angles is uniform and not related to the inclination. Because we observe that outflows are ubiquitous for $i \leq 45 \degree$, there can be no winds with half-opening angles less than $45 \degree$, else the outflow fraction would decrease before $45 \degree$. Thus, the differences in outflow fraction between adjacent inclination intervals when $i > 45 \degree$ represent the proportions of
outflows with inclinations in each interval. The predicted distribution of half-opening angles is shown in Figure~\ref{Fig:hist_opengingangle}. As expected, there are no galactic neutral gas outflows with half-opening angles less than $45 \degree$. 
Propagating the outflow fraction uncertainty to the half-opening angle distribution, the typical uncertainty in each bin is $\sim$0.1(10$\%$). Therefore, for the distribution of broader half-opening angles ($> 45 \degree$), we do not observe any significant patterns.

\section{Discussion}\label{sec.discussion}
\subsection{Outflow versus Star Formation Rate Evolution in Post-Starbursts}
\label{sec.discuss_vwind_age}
\begin{figure}
    \includegraphics[width=1\columnwidth]{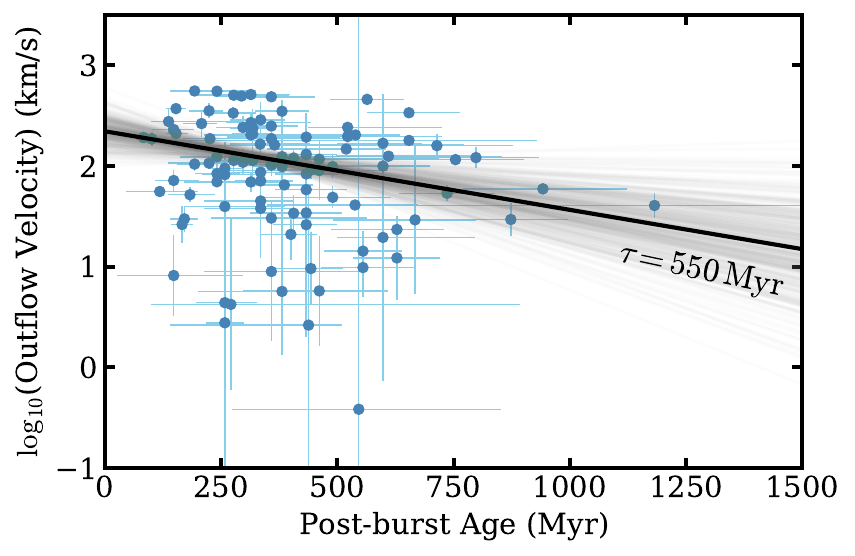}
\caption{Overall trend of outflow velocity with post-burst age for the outflow-detected post-starburst galaxy sample, without removing the stellar mass dependency. We fit these data using an exponential decline, finding the best-fit (black line) timescale over which the outflows slow is $550\pm200~\mathrm{Myr}$. The grey lines represent the dispersion of best-fit timescales. 
In the post-starbursts, the winds decline over a longer timescale than the SFR diminishes ($200\pm50~\mathrm{Myr}$; as shown in Figure~\ref{Fig:SFR_sequence}), suggesting that the outflows are decoupled from the current star formation rate and disfavoring the scenario where stars are responsible for the observed outflows.
}

\label{Fig:outflow_timescale}
\end{figure}

The results in Section~\ref{sec.vwind_age} show that winds are declining over the star formation declining sequence and with post-burst age. To further understand the connection between the evolution of outflows and star formation activity, i.e., how they affect each other, we fit the outflow and SFR trends for the post-starburst galaxy sample. Here we focus on the outflow-detected post-starburst sample, given that only outflows contribute to the expulsion of gas from the galaxy and could thereby reduce star formation. We fit the post-starburst age - $\ln(\text{outflow velocity})$ relation using {\tt\string PYTHON LINMIX} package\footnote{\url{https://github.com/jmeyers314/linmix}} in Figure~\ref{Fig:outflow_timescale}. The best-fit
exponential timescale for the wind decline is 550 Myr, with an uncertainty of about 200 Myr.

The best-fit timescale for the decline in SFR over the same period is 200 Myr, with an uncertainty of $\sim$50 Myr. To test the uncertainty introduced by the AGN contribution correction to the SFR measurements, we re-fit for the SFR timescale without the AGN correction
and obtain consistent results.
\citet{2023ApJ...942...25F} found that the decline in SFR for their post-starburst galaxies occurs over two different 
timescales: early (65 Myr) and late (480 Myr), with a turning point at 77 Myr. 
However, their sample included many galaxies younger than ours here---our youngest post-starburst galaxy with a SFR measurement 
is 83 Myr, older than 
\citet{2023ApJ...942...25F}'s
turning point age.  
Compared with the SFR decline timescale, the outflow velocity timescale is longer, suggesting that the outflows are de-coupled from the current (low levels of) star formation. This decoupling suggests that outflows in post-starburst galaxies are not driven primarily by stellar winds or supernovae.

\subsection{Fate of Post-Starburst Outflows}\label{sec.discuss_escape}

\begin{figure}
\centering
    \includegraphics[width=1.0\columnwidth]{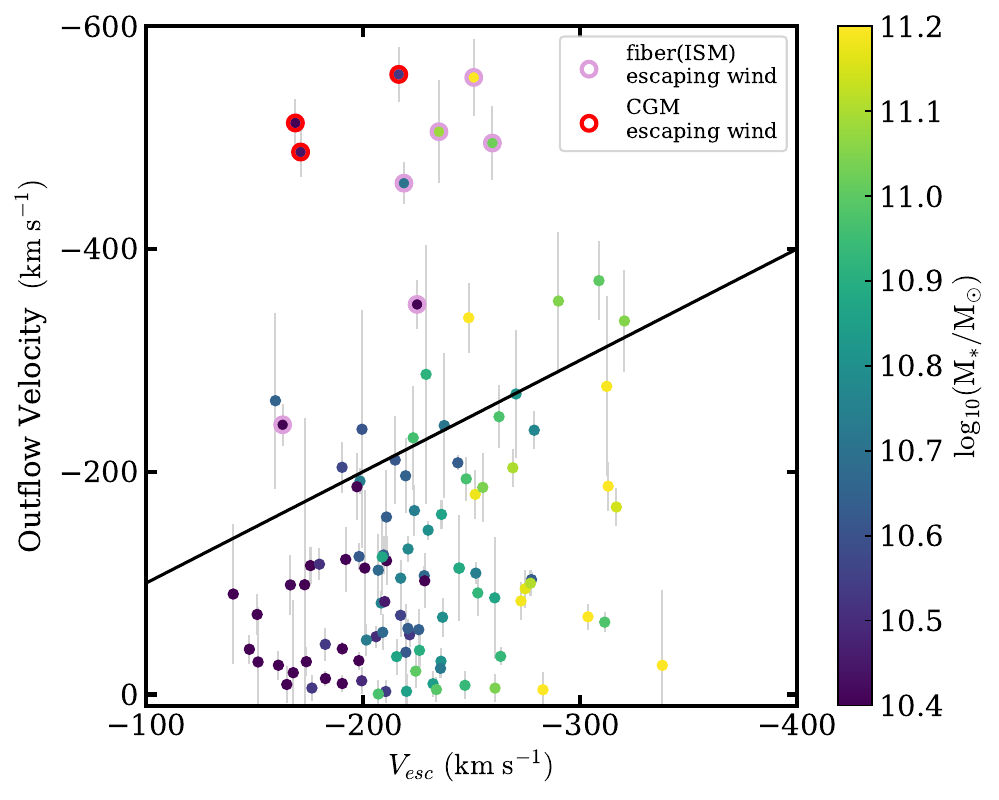}
\caption{Outflow velocities in post-starburst galaxies as a function of escape velocity within the SDSS fiber, whose aperture is comparable to the scale of ISM. The color-coding is by stellar mass. The black line represents the 1:1 relation. Nine of the 105 galaxies (pink or red circles) have outflow velocities 3$\sigma$ higher than the ISM escape velocity, and three (red circles) have outflow velocities  3$\sigma$ higher than the escape velocity for the CGM. Even considering that some of these sightline velocities are lower limits due to projection effects, statistically roughly $30\%$ of the outflows could exceed the ISM escape velocity, indicating that most are bound within the host galaxy.} 
\label{Fig:vel_escape}
\end{figure}

In this section, we test whether the $\nad$ outflows in the post-starburst galaxy sample have the capability of removing gas
from the interstellar medium of the galaxy, which is roughly the scale probed 
by the SDSS fiber over this redshift range. We compare wind velocity to ISM galaxy escape velocity in Figure~\ref{Fig:vel_escape}. 
The ISM escape velocity is estimated as $V_{\mathrm{esc}}\approx (2GM_{\mathrm{fib}}/r_{\mathrm{fib}})^{1/2} $, where $M_{\mathrm{fib}}$ is the stellar mass within the 3" SDSS fiber from the MPA-JHU catalog, and $R_{\mathrm{fib}}$ is the
physical scale of the fiber at the galaxy's redshift (0.3-7 kpc).
We separately estimate the escape velocity necessary to leave the galaxy's halo (i.e., the circumgalactic medium (CGM) escape velocity) as $V_{\mathrm{h,esc}}\approx3V_{\mathrm{h,vir}}$ (e.g., \citealt{2009ApJ...692..187W}), where $V_{\mathrm{h,vir}}$ is the halo virial velocity. We convert stellar mass into the halo mass using the relation from \citet{behroozi2019} and then calculate the halo virial velocity using the relation from \citet{2004MNRAS.355..694M}.

Only 9/105 ($9\%$) of post-starburst galaxies
have fast enough outflows to escape the ISM, i.e., have wind velocities 
3$\sigma$ higher than the ISM escape velocity. Only 3/9 ISM-escaping outflows can also escape from the halo.
To consider the sightline projection effects, which may cause us to underestimate the actual wind velocity,
we randomly select projected angles $\theta$ from $0\degree$ to $90\degree$ from a uniform distribution over 1000 trials. Then we perform a projection correction ($\Delta V_{real} =\Delta V/\cos{\theta}$). We re-calculate the percentage of ISM-escaping 
winds, finding $\sim$30$\%$, with an upper-limit of $50\%$.  This result confirms statistically
that most of the $\nad$ outflows detected in our post-starburst sample cannot expel gas from the galaxy. This gas would fall back into the disc, perhaps in  ``galactic fountains''.

Even if many winds cannot efficiently
remove gas, they might still affect any gas reservoirs within the galaxy. \citet{smercina_after_2022} found high internal turbulent pressure,
revealed by compact CO gas reservoirs and high-velocity dispersions, 
in six gas- and dust-rich post-starburst galaxies,
and \citet{2023ApJ...942...25F} detected low molecular gas excitation, suggesting that lower gas densities may suppress star formation during the post-starburst phase. Modest outflows within the ISM could both trigger high turbulence and prevent the cold gas from collapsing to form new stars. While these two phenomena were observed for molecular gas, our neutral outflows could also be associated with them, as multi-phase outflows are very common in post-starburst galaxies \citep{baron_multiphase_2021,luo_2022}.

\subsection{Origins of Post-Starburst Outflows}

\subsubsection{Associated with Star Formation}\label{sec.discuss_origin_SF}

As we mentioned in Section~\ref{sec.discuss_vwind_age}, the outflow velocities in post-starburst galaxies diminish over a timescale that is longer than that over which the SFR declines.
To further test if this de-coupling of winds and SFR disfavors a stellar origin for the winds, we compare the correlation between outflow velocity and SFR to the predictions of the simple analytic model for supernova explosion momentum-driven outflows introduced by \citet{2022ApJ...933..222X}. This model describes the behaviors of clouds under a combination of outward momentum from the starburst and the inward force of gravity. It also assumes that the gas we see is swept-up, ambient gas at the surface of the expanding wind bubble driven by the wind momentum flux (see Section 6.2 of \citealt{2022ApJ...933..222X} for details).
The predicted scaling relation between outflow velocity and SFR is then $\log(\text{outflow velocity})\propto 0.25\log(\text{SFR})$. \citet{2022ApJ...933..222X} detected 
ionized gas outflows in a sample of 45 low-redshift ($z<0.2$) star-forming galaxies; their best-fit slope to the ionized outflow velocity versus SFR relation (0.224 in the logarithmic scale) is consistent with the model.

\begin{figure}
    \includegraphics[width=1\columnwidth]{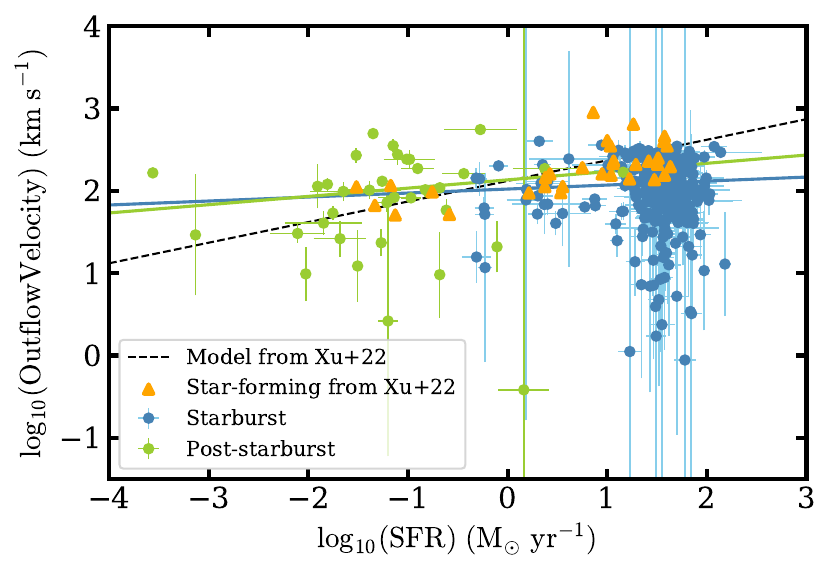}
\caption{Our neutral gas outflow velocities for starburst (blue) and post-starburst galaxies (green) as a function of the global H$\alpha$-based SFR. 
The best fit lines are shown for the starbursts ($\text{slope}=0.05 \pm 0.04$; blue) and post-starbursts ($\text{slope}=0.10 \pm 0.09$; green).
The prediction from the 
supernova explosion momentum flux-driven model of \citet{2022ApJ...933..222X} is $\log(\text{outflow velocity})\propto 0.25\log(\text{SFR})$ (black dashed line). The orange triangles represent star-forming galaxies with ionized outflow detections from \citet{2022ApJ...933..222X}, which are consistent with the model line. 
The model slope is 
somewhat steeper than the observed slopes for our starbursts and post-starbursts, which are consistent with zero. This apparent discrepancy suggests that neutral outflows have a weaker SFR dependency than ionized outflows.
Alternatively, 
the neutral winds may not be driven by supernovae.
For the starburst sample, there is other evidence (see Figure~\ref{Fig:trend_inclin}) implying that stellar winds do play a role.}
\label{Fig:SFR_outflowvel}
\end{figure}

If the detected winds in our post-starburst galaxies are driven by supernovae, and cold neutral clouds are coupled to hot ionized winds, we would expect a similar outflow velocity-SFR relation. However, Figure~\ref{Fig:SFR_outflowvel} shows that 
the model slope is somewhat steeper than for our starburst (slope $=0.05 \pm 0.04$) and post-starburst ($0.10 \pm 0.09$) samples, which have slopes consistent with zero.
The apparent discrepancy with the model might arise if the neutral phase of star formation-driven outflows has a weaker SFR dependence than the ionized phase. 
Another possibility is that the outflows arise mostly from a mechanism other than supernovae.
At least for the starburst galaxies,  
the inclination dependence discussed in Section~\ref{sec.inclin_result} still implies
that star formation-driven winds play a role.

There is another interpretation to consider.
Even if the outflows detected in post-starbursts are unrelated to current star formation, the cool neutral gas in the flow may be long-lived, representing 
a relic outflow (e.g., \citealt{2018MNRAS.481.1873L}) launched near the peak starburst time that has lingered in the ISM. In this scenario, the covering factor of the outflowing cold clouds as seen by the galaxy decreases in time.
However, we find that $\ewexc$, which is proportional to the covering factor of the neutral gas, is not correlated with post-burst age for our sample, arguing against this interpretation.

\subsubsection{Associated with Nuclear Activity}\label{sec.discuss_origin_agn}

\begin{figure}
\centering
    \includegraphics[width=0.95\columnwidth]{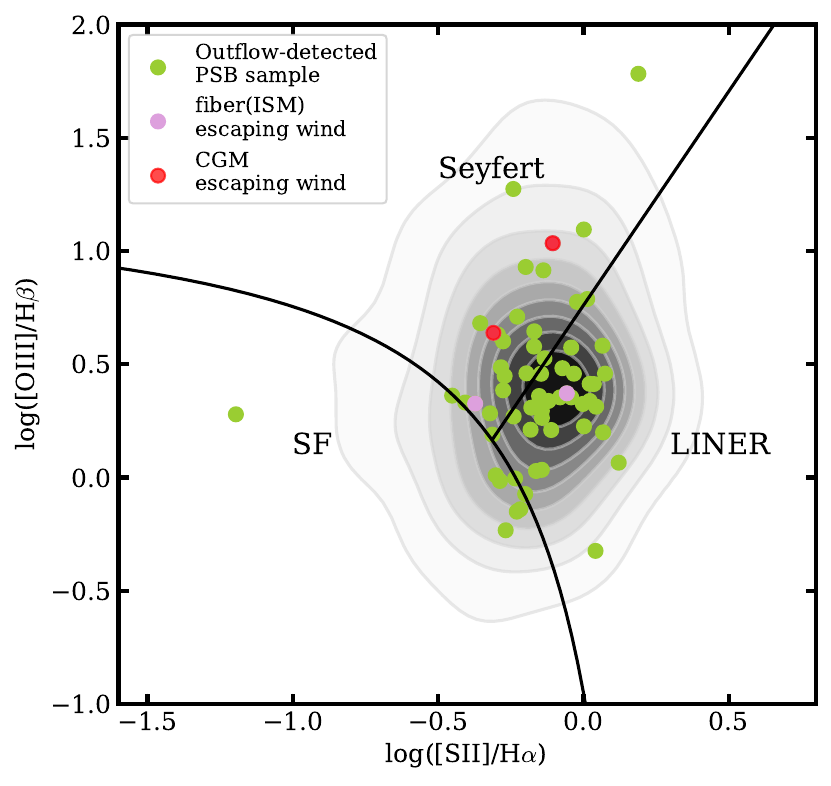}
\caption{Post-starburst galaxies with outflows on the $[\SII]$-BPT diagram \citep{veilleux_spectral_1987}. The gray contours are for all post-starburst galaxies in our sample. The
outflow-detected subsample is plotted as green points. Three of the four ($75\%$) galaxies with ISM-escaping outflows (pink or red points) lie in the Seyfert part of the diagram, including both galaxies with CGM-escaping outflows (red points). Compared to the fraction of all outflow-detected post-starbursts that are Seyferts (24/61, $39\%$)
and the fraction of all post-starbursts that are Seyferts (124/336, $37\%$), the fraction of post-starbursts with escaping outflows that are Seyferts is higher.
}
\label{Fig:BPT}
\end{figure}

AGN are likely to contribute to the outflows seen in post-starburst galaxies. \citet{baron_multiphase_2021} detected neutral outflows in 40\% of their 144 post-starburst galaxies with AGN and ionized outflows. The median velocity of neutral outflows in their post-starburst galaxies is $-633~\kms$, much higher than the mean $\Delta V$ of our post-starburst sample ($-71.4~\kms$). This difference is likely due to our selection against significant AGN spectral features
and to our averaging over all bulk motions, including inflows. 
If the difference arises from sample selection, it suggests that
strong AGN (included in the \citet{baron_multiphase_2021} sample) produce faster outflows than those driven by weaker AGN or by stellar winds alone (our sample).

Even though our post-starburst sample is biased against \textbf{strong} AGN, nuclear activity may have been stronger at earlier times.
\citet{wild_timing_2010} explored the growth of black holes in 400 local galactic bulges that have had a starburst in the past 600 Myr and found that the AGN activity peaks 250 Myr after the starburst\footnote{Note that ``the time after starburst'' in \citet{wild_timing_2010} is the time since the starburst began,
not our ``post-burst age,'' which is the time since the starburst ended. As a result, ``250 Myr after the starburst'' in their paper is an earlier time than a post-burst age of 250 Myr, given that the typical starburst duration is about 100 Myr.}. On the simulation side, \citet{2023MNRAS.519.4966B} used IllustrisTNG to test the supermassive black hole (SMBH) accretion rate and SFR enhancement in post-merger galaxies---post-starburst galaxies often result from mergers---and found that accretion rate enhancements persist for up to 2 Gyr after coalescence. 
\citet{2012MNRAS.420..878K}
found that the timescale for an AGN to vary from quasar-like to LINER-like is much faster than that of the post-starburst phase. Therefore, while the relative timescales among the starburst,  nuclear activity, and outflows are not well-constrained, 
outflows in post-starburst galaxies could be the relics of AGN in the starburst or early post-starburst phase. 

To further test the hypothesis that AGN is responsible for the post-starburst winds, we place our post-starburst galaxies on the $[\SII]$-BPT diagram \citep{veilleux_spectral_1987} in Figure~\ref{Fig:BPT} \footnote{We use the $[\SII]$-BPT diagram, because it separates the ``Seyfert'' and ``LINER'' regions better than the $[\NII]$ version.}.The BPT emission lines come from the MPA-JHU catalogs \citep{kauffmann_stellar_2003,brinchmann_physical_2004,tremonti_origin_2004}.
Of the 516 post-starburst galaxies, 336 have positive BPT emission line measurements from the MPA-JHU catalogs, including 61 outflow-detected galaxies.

Although we select against galaxies with strong emission lines, 
many post-starburst galaxies in our sample still have weak AGN or LINER features after stellar spectral continuum subtraction. Given the high Seyfert/LINER fraction of outflow-detected galaxies (53/61, $87\%$), it is natural to assume that the AGN and outflows are connected. However, to fully test this point requires that we check whether the chance of a Seyfert/LINER post-starbursts having wind is significantly higher than expected from the fraction of all post-starbursts with winds (whether we can place them on the BPT diagram or not). For our sample, this assumption yields that the outflow detection rate in Seyfert/LINER post-starbursts is not statistically higher than the rate in all post-starbursts; both rates are roughly $20\%$. So, the high AGN and high outflow fractions in our sample do not in themselves lend support to the picture of AGN-driven winds.

If we instead focus on the four with ISM-escaping outflows
(see Section~\ref{sec.discuss_escape}), 
we find that three lie in the Seyfert region, including both galaxies with CGM-escaping outflows. (The remaining galaxy with an ISM-escaping outflow is a LINER.) Compared to the Seyfert-only fractions of the outflow-detected post-starburst sample (24/61, $39\%$) and of the entire post-starburst sample (124/336, $37\%$), the Seyfert fraction of post-starbursts with ISM-escaping outflows ($75\%$) is higher. Together with the observed differences in outflow velocity between the \citet{baron_multiphase_2021} sample and ours, this result, while not statistically significant, suggests that AGN could drive the most effective outflows in post-starburst galaxies.

\begin{figure}
\centering
    \includegraphics[width=1.0\columnwidth]{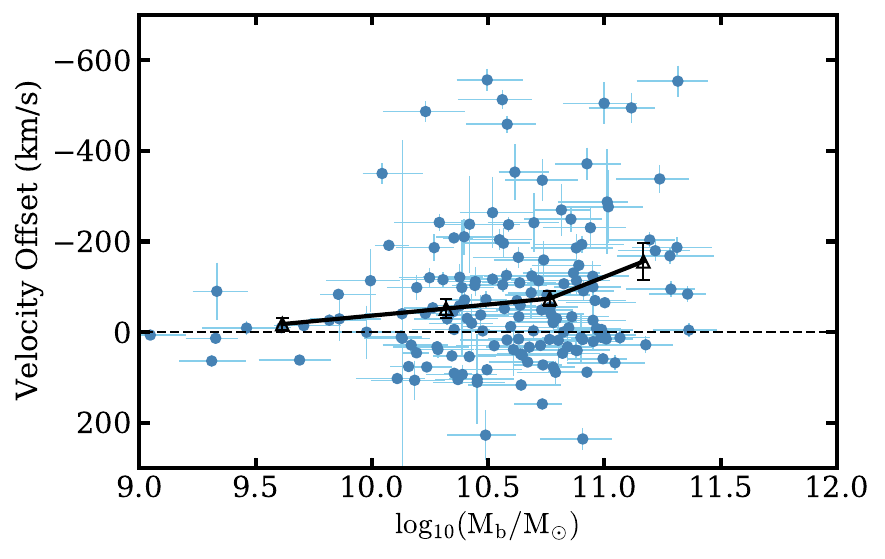}
\caption{Correlation between velocity offset $\Delta V$ and bulge mass $M_{bulge}$ for the post-starburst sample.  The Spearman correlation coefficient is $\rho=-0.14,~1.78\sigma$. The black triangles show the mean velocity offset in each bulge mass bin. 
The $\Delta V$-$M_{bulge}$ relation here is similar to the $\Delta V$-$M_{*}$ relation in the top left panel of Figure~\ref{Fig:original_vel_trend}, suggesting that the latter relation is tied to bulge mass and thus black hole mass.}
\label{Fig:vel_bulgemass}
\end{figure}

Further indirect evidence for AGN-driven outflows in post-starbursts is from the correlation between $\Delta V$ and $M_{*}$ (Section~\ref{sec.vwind_age}). This correlation may be a secondary correlation caused by the real $\Delta V$-post-burst age relation (panel (c) of Figure~\ref{Fig:original_vel_trend}) and $M_{*}$-post-burst age relation (panel (a) of Figure~\ref{Fig:original_trend_mstar_psbage_z}).
It is also possible that the $\Delta V$-$M_{*}$ relation is driven by trends with black hole mass, i.e., more massive galaxies have more massive black holes (BHs) that drive faster outflows. To test this second possibility, we check whether $\Delta V$ correlates with bulge mass ($M_{bulge}$) as it does with stellar mass. We obtain $M_{bulge}$ values from \citet{mendel_catalog_2014}. 
Figure~\ref{Fig:vel_bulgemass} shows that the $\Delta V$-$M_{bulge}$ correlation has similar strength and significance  ($\rho=-0.14,~1.78\sigma$) to the $\Delta V$-$M_{*}$ relation (panel (a) in Figure~\ref{Fig:original_vel_trend}). Given that bulge mass is strongly correlated with black hole mass ($M_{BH}$), the observed $\Delta V$-$M_{*}$ relation could be caused by the $\Delta V$-$M_{BH}$ relation, a finding consistent with the AGN-driven outflow scenario. 

In addition to the possibility of AGN outflows, the high tidal disruption event (TDE) rate in post-starburst galaxies \citep{2016ApJ...818L..21F,2020SSRv..216...32F} and the likely resulting energy input \citep{2021ApJ...906..101M} could generate low-level nuclear winds that contribute to the outflows we see here. \citet{smercina_after_2022} makes a similar argument for TDEs as the source of the turbulent molecular gas reservoirs they observed in post-starburst galaxies.

\section{Conclusions}\label{sec.concl}

In this paper, we use the $\nad$ absorption doublet to analyze bulk neutral gas motions in the ISM of $\sim$80,000 $0.010 < z < 0.325$ galaxies from the SDSS DR12 survey. 
We first construct a galaxy evolutionary sequence in declining SFR, from
starburst to post-starburst to quiescence.
To construct this sequence,
we find likely progenitors (starburst galaxies) and descendants (quiescent galaxies) of the post-starburst galaxies in F18 by constraining their stellar masses and star formation rates.
To isolate any gas flow contribution, we subtract the best-fit stellar continuum from the spectrum to obtain a residual $\nad$ line that arises from the ISM. We then measure the positive (inflow) or negative (outflow) velocity shift $\Delta V$ of this line relative to the systemic velocity of the stars. 

Our main findings are:

\begin{enumerate}

    \item Bulk flows---in or out of the galaxy---are detected along the entire evolutionary sequence, especially at higher host stellar masses, i.e., $\lgmstar>10$. Both the starburst and post-starburst samples have
    $\Delta V$ distributions with statistically significant 
    negative tails, implying that we have detected many real outflows.
    The quiescent sample has a positive $\Delta V$ tail that reveals significant inflows. These inflows are associated with smaller equivalent width residual $\nad$ lines than the starburst and post-starburst outflows, indicating lower mass loading;
    this result may explain why, despite inflows being common in our "red, dead" galaxies, they do not trigger significant star formation.
    
    \item We see further evidence for outflows in the $\nad$ line profiles.
    Among those hosts with bulk flows, 
    9/370 (2.4\%) of starbursts, 11/163  (6.7\%) of post-starbursts, and 13/4598 (0.3\%) of quiescents have P-Cygni profiles that indicate outflows.  
    As a consistency check on our fitting methodology, we confirm that all of these systems are classified as having outflows, i.e., negative velocity offsets.
    
     \item For all three galaxy samples, the mean outflow velocities at high $M_{*}$ ($\lgmstar\geq 11$) are faster than at low $M_{*}$ ($\lgmstar<11$).  For the post-starbursts, outflow velocity increases significantly with $M_{*}$. It is possible that this trend is tied instead to bulge mass and thus black hole mass, as we find a similarly strong $\Delta V$-$M_{bulge}$ relation for post-starbursts and independent bulge mass estimates.
    
    \item Outflows diminish as 
    galaxies age and star formation declines.
    The fraction of hosts with outflows decreases along the galaxy evolution sequence from starbursts ($76\pm2\%$) to post-starbursts ($64\pm6\%$) to quiescents ($33\%$, 3$\sigma$ upper limit). Similarly, the mean velocity offset of the gas changes from $-84.6\pm5.9$ to $-71.4\pm11.5$ to $76.6\pm2.3$, i.e., 
    from outflowing to inflowing.
    Even within the post-starburst sample, and after controlling for host stellar mass, the mean flow speed decreases with time elapsed since the starburst ended.

    \item For the post-starburst sample, the SFR declines more quickly with post-burst age than the outflow velocity does, suggesting that stellar winds are not the principal driver of the outflows.

    \item For the post-starbursts, only a small fraction of measured outflows significantly exceed the escape velocity of the interstellar medium defined by the SDSS fiber aperture (9/105) or the circum-galactic medium defined by virial radius of the dark matter halo (3/105). 
    Although we measure only the line-of-sight component of the wind, most ($\sim70\%$)
    winds are still statistically likely to remain within the host galaxy.
    
    \item For the post-starbursts, 3/4 (75\%)
    with ISM-escaping outflows, including both of the galaxies with CGM-escaping outflows, that can be placed on the BPT diagram lie in the Seyfert region (despite our post-starburst sample selection against H$\alpha$ emission). 
    In comparison, the Seyfert fraction of the entire post-starburst sample is 124/336 (37\%) and that of the subsample with outflows of any kind is 24/61 (39\%). 
    
    \item For the starburst galaxies with discs, the outflows are dependent on disc inclination, which favors a wind model in which gas is ejected perpendicularly from the disc plane. The wind velocity decreases as the disc becomes more edge-on, and the outflow fraction, constant at $\sim$90$\%$ for disc inclinations $i<45\degree$,
    steadily decreases from $\sim$90$\%$ to 20$\%$
    for $i>45\degree$. Our analysis limits the outflow 1/2-opening angle to roughly $ > 45\degree$.
    
\end{enumerate}

There are conflicting clues as to the nature of the winds. 
For the post-starburst galaxies, the high Seyfert fraction associated with the strongest outflows suggests an AGN connection to gas removal from these galaxies. Another point favoring a nuclear over stellar origin for post-starburst outflows is the different timescales over which the SFR and wind speed decline. The post-starburst $\Delta V$-$M_{bulge}$ relation observed here is also consistent with the AGN-driven picture if $M_{bulge}$ is a proxy for black hole mass. Furthermore, the $\Delta V$-SFR relations for starbursts and post-starbursts are marginally inconsistent with the prediction of a simple supernovae momentum-driven wind model.
On the other hand, for starburst galaxies with discs, the trends of inclination versus $\Delta V$ and outflow fraction support a perpendicular wind model, a geometry most likely to arise from star formation throughout the disc.

Critically, the existence of outflows does not demonstrate ``feedback,'' as it is not yet clear how (or even if) these winds regulate star formation in their hosts. The intriguing drop observed here in outflow speed and outflow fraction with SFR along the starburst/ post-starburst/quiescent sequence does not imply causation. Given their slow speeds, many of these outflows are likely
to remain bound to their host, unable to remove its gas. Yet, outflows may reduce the SFR by generating turbulence or re-distributing the ISM.
 
To better understand the whole picture of outflows---how they affect and are affected by star formation history---a similar analysis is required \emph{within} individual galaxies. Spatially-resolved spectroscopy, e.g., from the SAMI IFS data \citep{2021MNRAS.505..991C} and the MaNGA IFU data \citep{2015ApJ...798....7B}, across the galaxy evolution sequence defined here
could reveal the source of the outflows
by localizing them to the disc or to the galactic center.
Resolving the spatial extent of the outflows will also constrain the mass/energy loss \citep{2020MNRAS.493.3081R, 2021MNRAS.503.5134A, 2022MNRAS.511.4223A}, quantifying the winds' impact on the ISM and CGM, a true test of feedback processes.

\section*{Acknowledgements}
We thank the anonymous referee for a thoughtful report that improved this paper. We also thank George Rieke and Dennis Zaritsky for helpful discussions.
YS and AIZ acknowledge support from NASA ADAP grant \#80NSSC21K0988.
AIZ also thanks the hospitality of the Columbia Astrophysics Laboratory at Columbia University, where some of this work was completed.
YY is supported by the Basic Science Research Program through the National Research Foundation of Korea funded by the Ministry of Science, ICT \& Future Planning (2019R1A2C4069803).

\section*{Data Availability}

 The data underlying this article were accessed from the publicly available Data Release 12 (DR12; \citealt{alam_eleventh_2015}) of the Sloan Digital Sky Survey (SDSS-III; \citealt{eisenstein_sdss-iii_2011}). Specifically, we make use of the SDSS optical spectra. See the SDSS web page (\url{https://dr12.sdss.org/}) for more details of the DR12 spectra. All spectra can be accessed using the traditional SDSS Science Archive Server (SAS) at \url{https://data.sdss.org/sas/dr12/sdss/spectro/}.

 We have tabulated the measured properties of our starburst, post-starburst, and quiescent galaxy samples, including the $\nad$ equivalent width and velocity offsets derived from the optical spectra. We also include the inclinations of starburst galaxies and H$\alpha$-based SFR measurements for the post-starburst and quiescent galaxies. We also tabulate the redshifts, stellar masses, and SFRs of starburst galaxies from the MPA-JHU catalogs, and the post-burst ages and burst mass fractions of the post-starbursts from \citet{french_clocking_2018}. These data will be made publicly available through the online supplementary material upon the acceptance of the manuscript.



\bibliographystyle{mnras}
\bibliography{reference-final} 







\bsp	
\label{lastpage}
\end{document}